\documentclass[12pt]{article}
\pdfoutput=1
\usepackage{epsf,amsfonts,amssymb,epsfig,amsmath,mathtools,graphics,slashed}
\usepackage{hyperref,hep,graphicx,subfig}
\usepackage{rotating}
\newcommand{\nn}{\notag \\}

\makeatletter

\@addtoreset{equation}{section}
\makeatother

\begin{document}

\begin{titlepage}

\vfill

\begin{flushright}
Imperial/TP/2017/JG/03\\
DCPT-17/09
\end{flushright}

\vfill

\begin{center}
   \baselineskip=16pt
   {\Large\bf Boomerang RG flows in M-theory\\ with intermediate scaling}
  \vskip 1.5cm
  \vskip 1.5cm
Aristomenis Donos$^1$, Jerome P. Gauntlett$^2$\\ Christopher Rosen$^2$ and Omar Sosa-Rodriguez$^1$\\
     \vskip .6cm
     \begin{small}
      \textit{$^1$Centre for Particle Theory and Department of Mathematical Sciences\\Durham University, Durham, DH1 3LE, U.K.}
        \end{small}\\    
         \begin{small}\vskip .6cm
      \textit{$^2$Blackett Laboratory, 
  Imperial College\\ London, SW7 2AZ, U.K.}
        \end{small}\\
        \end{center}
     \vskip .6cm
\vfill

\begin{center}
\textbf{Abstract}
\end{center}
\begin{quote}
We construct novel RG flows of D=11 supergravity that asymptotically approach 
$AdS_4\times S^7$ in the UV with deformations that break spatial translations
in the dual field theory.
In the IR the solutions return to exactly the same $AdS_4\times S^7$ vacuum,
with a renormalisation of relative length scales, 
and hence we refer to the flows as `boomerang RG flows'.
For sufficiently large deformations, on the way to the IR the solutions also
approach two distinct intermediate scaling regimes, each with hyperscaling violation. 
The first regime is Lorentz invariant with dynamical exponent $z=1$ 
while the second has $z=5/2$. Neither of the two intermediate scaling regimes are associated with exact hyperscaling violation
solutions of $D=11$ supergravity. 
The RG flow solutions are constructed using the four dimensional $N=2$ STU gauged supergravity theory with vanishing gauge fields, but non-vanishing scalar and pseudoscalar fields. In the ABJM dual field theory the flows are driven by spatially modulated deformation parameters
for scalar and fermion bilinear operators.

\end{quote}

\vfill

\end{titlepage}

\setcounter{equation}{0}
\section{Introduction}
The AdS/CFT correspondence provides us with powerful tools to investigate the behaviour of strongly coupled 
conformal field theories that have been deformed by operators that explicitly break spatial translations. Indeed,
by solving suitable gravitational equations we can study how such systems evolve under the renormalisation group as well as
study their properties at finite temperature.
One motivation for these investigations is that they
provide a framework for studying strongly coupled systems with novel thermoelectric transport properties, which could connect
with real systems seen in Nature.

Generically, the construction of the relevant RG flows requires solving a system of partial differential equations since the bulk fields will depend on both the spatial directions of the field theory
as well as a holographic radial direction.
An interesting exception is provided by the Q-lattice constructions \cite{Donos:2013eha}. These constructions require the existence of a 
global symmetry in the bulk which is then used to develop an ansatz for the bulk fields in which the dependence on the spatial
directions is solved exactly. This leads to a system of ordinary differential equations for a set of functions that
just depend on the holographic radial coordinate. It is of particular interest to look for Q-lattice constructions in a top-down context.

In type IIB supergravity there is a rich class of examples associated with the family of $AdS_5\times X_5$ vacuum solutions, 
where $X_5$ is an Einstein space \cite{Donos:2016zpf}. The solutions can be constructed using a $D=5$ theory of gravity which is obtained
as a consistent Kaluza-Klein reduction of type IIB supergravity on $X_5$. The $D=5$ theory couples the metric to a complex scalar field $\tau$, which
incorporates the axion and dilaton of type IIB supergravity, and there is a natural global $SL(2)$ symmetry acting on $\tau$. 

For each of the three conjugacy classes of $SL(2)$ there is an associated 
Q-lattice construction of type IIB supergravity
which is spatially anisotropic, breaking translations in one of the three spatial directions. 
The hyperbolic conjugacy class corresponds to a linear dilaton solution in which the dilaton depends linearly on the preferred spatial direction 
\cite{Jain:2014vka} and the axion is trivial. The parabolic conjugacy class corresponds to
a linear axion solution, with non-trivial dilaton, \cite{Azeyanagi:2009pr,Mateos:2011ix,Mateos:2011tv} which flows 
to a lifshitz-like fixed point\footnote{For the special case when $X^5=S^5$, associated with $N=4$ super Yang-Mills theory, this fixed point is unstable
\cite{Azeyanagi:2009pr} and there is a more elaborate phase diagram which has been partially explored in \cite{Banks:2015aca,Banks:2016fab}.} 
in the IR. 

Finally, there is a Q-lattice construction of type IIB supergravity associated with the elliptic conjugacy class which, unlike the other two classes, breaks translations in a periodic way \cite{Donos:2016zpf}. The elliptic solution flows from $AdS_5$ in the UV to the {\it same} $AdS_5$ solution 
in the IR. While the central charge of the CFT has the same value in the UV and the IR, there is
a renormalisation of relative length scales. We will refer to such flows as  `boomerang RG flows' and we note that they have also been seen in other constructions with broken translations \cite{Chesler:2013qla,Donos:2014gya}. Furthermore, for sufficiently large deformations, the boomerang RG flow has an intermediate scaling regime that is dominated by the Lifshitz-like fixed point of \cite{Azeyanagi:2009pr} that appears in the parabolic class. Such flows can therefore be viewed as an interesting
framework for resolving the singularity of the Lifshitz-like fixed point, differing from other
singularity resolving flows \cite{Harrison:2012vy,Bao:2012yt,Bhattacharya:2012zu,Kundu:2012jn,Donos:2012yi,Bhattacharya:2014dea}.

Motivated by these type IIB constructions, in this paper we will construct examples of Q-lattices of $D=11$ supergravity which 
are associated with the $AdS_4\times S^7$ vacuum solution. While we find some similarities with the type IIB solutions we also find
many new features. The new constructions
will be made in the $N=2$ STU gauged supergravity theory in four dimensions \cite{Cvetic:1999xp}. 
Recall that this theory arises from a
consistent truncation of $N=8$ gauged supergravity and hence any solution can be uplifted on the seven sphere, or a quotient thereof, 
to obtain a solution of $D=11$ supergravity \cite{Cvetic:1999xp,Cvetic:2000tb,Azizi:2016noi}. 
As such our solutions are directly relevant to ABJM theory \cite{Aharony:2008ug}.

The new $D=4$ solutions involve two complex scalar fields each of which parametrises $SL(2,R)/SO(2)$. The Lagrangian
has a potential term which breaks the $SL(2,R)$ symmetry down to $SO(2)$ and we use the latter for our Q-lattice construction.
When expanded about the $AdS_4$ vacuum these scalar fields are dual to relevant operators in the dual CFT;
this can be contrasted with the type IIB axion-dilaton which is massless and dual to a marginal complex operator 
(for any choice of $X_5$). 
Within the ABJM theory the scalar fields are dual to certain scalar and fermion bilinear operators and the RG flows are thus
being driven by spatially modulated mass deformations.

An additional difference with the type IIB flows is that having two complex scalars allows
us to break translation invariance in both spatial directions. Furthermore, this is achieved with a bulk metric that preserves spatial isotropy in the field theory directions.
We will construct a one parameter family of solutions, parametrised by the dimensionless ratio $\Gamma/k$, where $\Gamma$ governs the strength of the deformation of the relevant operators in the UV and $k$ is the wave number of the periodic spatial modulation.

\begin{figure}[t!]
\centering
\includegraphics[width=0.5\textwidth]{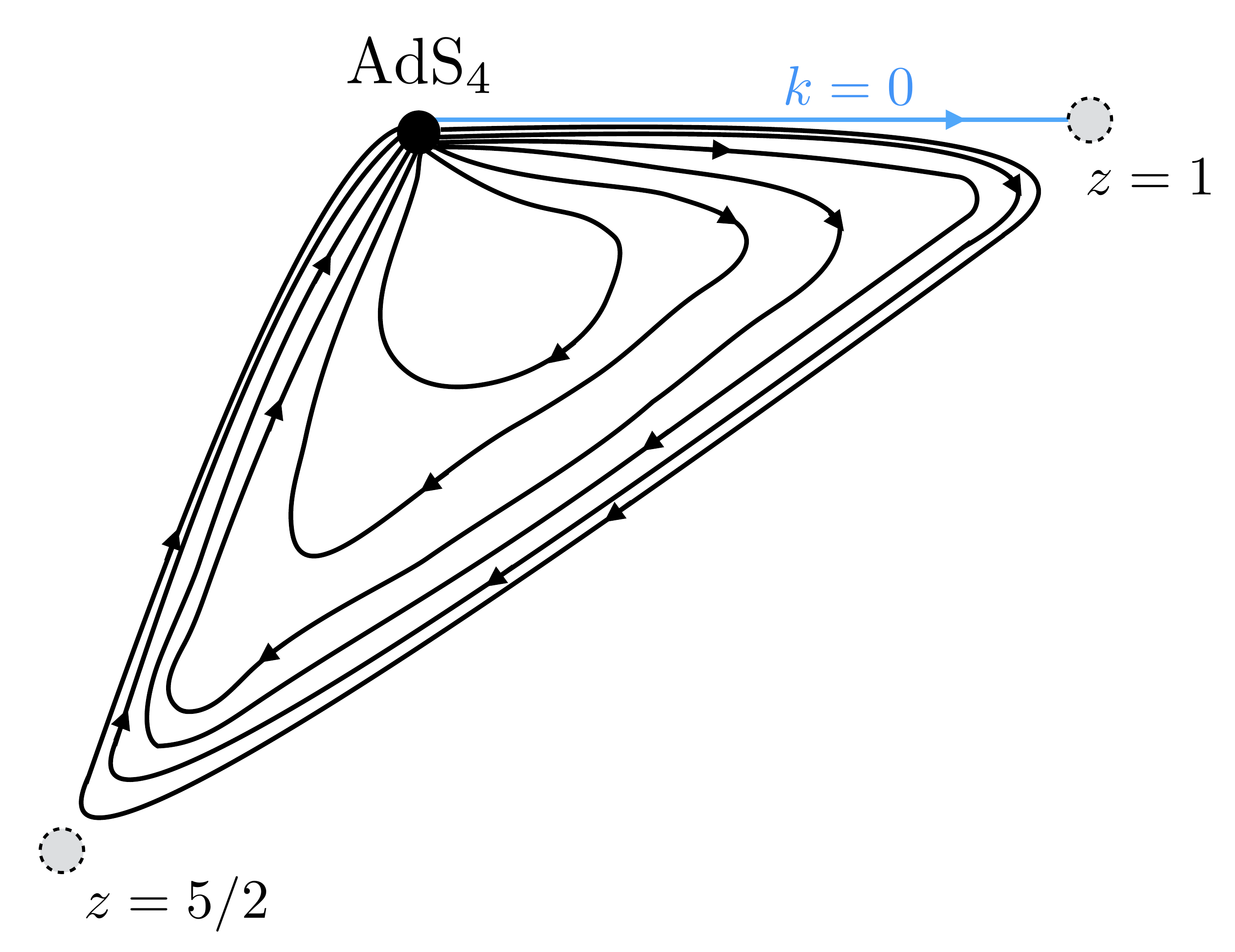}
\caption{Schematic picture of the family of boomerang RG flows, parametrised by $\Gamma/k$ which fixes the strength of the relevant UV deformation. They all flow from the $AdS_4$ vacuum in the UV to the same $AdS_4$ vacuum in the IR. For sufficiently large 
$\Gamma/k$ the solutions exhibit two intermediate scaling regimes in the bulk geometry. Both regions are of hyperscaling
violation form given in \eqref{genmet}: the first is Lorentz invariant with $z=1$, $\theta=-2$ while the second has $z=5/2$, $\theta=1$.
Neither of the two intermediate scaling regimes are associated with exact solutions of $D=11$ supergravity.
The blue `$k=0$' line is associated with a Lorentz invariant RG flow from $AdS_4$ in the UV to approximate
hyperscaling behaviour with $z=1$, $\theta=-2$ in the far IR.
\label{fig1}}
\end{figure}

Similar to the type IIB solutions, the new RG solutions are again boomerang flows, flowing from the $AdS_4$ vacuum solution
in the UV down to the same $AdS_4$ solution in the IR with renormalised relative length scales. In addition, for large enough values of $\Gamma/k$ we find that on the way to IR the RG flow exhibits intermediate scaling behaviour,
similar to the $D=5$ flows in the elliptic class. Interestingly, however, in contrast to the $D=5$ flows there are now two distinct intermediate scaling regimes and both exhibit hyperscaling violation \cite{Charmousis:2010zz,Ogawa:2011bz,Huijse:2011ef}. 
The first regime is Lorentz invariant with dynamical exponent $z=1$ and hyperscaling violation exponent $\theta=-2$, while the second has $z=5/2$ and hyperscaling violation exponent $\theta=1$. 
A schematic picture of the RG flows is presented in figure \ref{fig1}.

This first intermediate scaling regime is directly related to the fact that we are deforming by a relevant operator. Indeed, the dimensionless
deformation parameter, $\Gamma/k$, necessarily involves $k$ and hence one can anticipate that the $\Gamma/k\to\infty$ behaviour 
should approach that of RG flows with $k=0$ and $\Gamma\ne 0$. This simple observation indicates that
Poincar\'e invariant intermediate scaling will be a more general phenomena in systems with deformations of 
relevant operators that break translations. Note that it did not occur in the type IIB flows \cite{Donos:2016zpf} because the deformations by the axion and dilaton are associated with marginal operators.
The existence of the second intermediate scaling regime is less obvious, {\it a priori}, and furthermore it is an interesting fact that this regime appears for the same values of $\Gamma/k$ for which the first intermediate scaling appears.

Another difference with the $D=5$ flows, is that neither of the intermediate scaling behaviours are 
governed by fixed point solutions of the $D=4$ gauged supergravity theory. Indeed the fact that there is hyperscaling violation means that there is
a scalar field that is still running and hence they cannot be fixed point solutions. In fact, there are no exact hyperscaling violation solutions to
the equations of motion which are determining the scaling behaviour. 
To elucidate the first regime, we construct a Poincar\'e invariant RG flow with $\Gamma\ne0$ and $k=0$, which flows from $AdS_4$ in the UV and approaches a
hyperscaling violation behaviour in the far IR, without the far IR behaviour being itself a solution to the equations of motion.
It is the far IR scaling behaviour of this RG flow, which we call the `$k=0$ flow',
that governs the first intermediate scaling of the RG flows with broken translation invariance shown in figure \ref{fig1}. 

To understand the second scaling regime, we show that there is a hyperscaling violation solution with broken translation invariance and $z=5/2$, $\theta=1$
of an {\it auxiliary} theory of gravity, which has equations of motion that agree with the STU theory for large values of the modulus of the complex scalar fields. It is this 
solution which governs the second intermediate scaling of our RG flows shown in figure \ref{fig1}. 

We are unaware of other RG flows in holography which exhibit such novel intermediate scaling behaviour and anticipate that these, or closely related flows, will have interesting applications. It is worth highlighting that
using an auxiliary theory of gravity to govern intermediate scaling is rather simple and natural from the gravity side, but it is less so from the field theory point of view.  Roughly speaking, it is associated with moving to the boundary in the space of coupling constants.

We have organised the paper as follows. In section \ref{setup} we introduce the $D=4$ theory of gravity that we will study, as well as
the ansatz for the new RG flow solutions. In section \ref{intscal} we pause to discuss both the $k=0$ RG flow and also the 
scaling solution of an auxiliary theory of gravity, each of which governs an intermediate scaling behaviour of the boomerang RG flows. The main
results for the RG flows with intermediate scaling are presented in section \ref{rgflows}. In this section we also
discuss how the intermediate scaling manifests itself in the behaviour of thermodynamic quantities at finite temperature,
as well as in the behaviour of spectral functions of certain operators in the dual CFT at zero temperature.
Using a matched asymptotics argument\footnote{Matching arguments were also discussed in the context of intermediate scaling
of the optical conductivity in \cite{Bhattacharya:2014dea}.}, which provides sufficient conditions for the appearance of intermediate scaling behaviour,
 we will expose an interesting type of universality whereby scalar operators with different scaling dimensions, $\Delta$, in the dual CFT, up to some maximum value set by the details of the flow,
can have spectral functions with the same intermediate scaling behaviour for a certain range of frequency.
We conclude with some final comments in section \ref{fincom}. In appendix \ref{appa} we discuss the $D=4$ STU theory and also present an
ansatz that can be used to construct charged anisotropic Q-lattice solutions. 

\section{The set-up}\label{setup}
Our starting point is the $N=2$ truncation of maximal $N=8$ $SO(8)$ gauged supergravity in four dimensions, 
whose bosonic field content
consist of the metric, four $U(1)$ gauge-fields and three neutral complex scalar fields $\Phi_i$ which we write as
\begin{align}\label{xandyo}
\Phi_i=X_i+iY_i=\lambda_ie^{i\sigma_i}\,.
\end{align}
The Lagrangian is given in appendix \ref{appa}. The $X_i$ are components of the 35 scalars and the
$Y_i$ are components of the 35 pseudoscalars transforming in the ${\bf 35_v}$ and ${\bf 35_c}$ of the $SO(8)$ global symmetry of the $N=8$ theory, respectively \cite{Duff:1999gh}. 
Using the formula given in \cite{Cvetic:1999xp,Cvetic:2000tb,Azizi:2016noi}, any solution of the $N=2$ theory can be explicitly uplifted on a seven sphere to obtain an exact solution of $D=11$ supergravity. 

In the bulk of this paper we will be interested in solutions with vanishing gauge-fields and 
we will also truncate $\lambda_1=\sigma_1=0$, which we can do consistently. Thus, we consider the Lagrangian
\begin{equation}\label{eq:tLag2}
\mathcal{L} = R-\frac{1}{2}\sum_{i=2}^3\left[(\partial\lambda_i)^2+\sinh^2\lambda_i(\partial\sigma_i)^2\right]
+2(1+\cosh\lambda_2+\cosh\lambda_3)\,.
\end{equation}
The $AdS_4$ vacuum solution, with unit radius and $\lambda_2=\lambda_3=\sigma_2=\sigma_3=0$, uplifts to the maximally supersymmetric $AdS_4\times S^7$ solution.

We now introduce the following ansatz, which breaks translation invariance in both spatial directions $(x,y)$ of the dual field theory:
\begin{align}\label{ansatz}
\mathrm{d}s^2 &= -U(r)\mathrm{d}t^2+U(r)^{-1}{\mathrm{d}r^2}+e^{2V(r)}(\mathrm{d}x^2+\mathrm{d}y^2)\,,\nn
\sigma_2&=kx,\quad\sigma_3=ky,\qquad \lambda_2=\lambda_3=\gamma(r)\,.
\end{align}
%This ansatz depends on three functions $U,V$ and $\gamma$ of the radial coordinate $r$. 
Notice that the ansatz for the metric preserves the spatial isotropy in the $(x,y)$ directions.
Also, since the $\sigma_i$ are periodic variables, the dependence on the spatial coordinates is periodic in $x,y$ with period $2\pi/k$. This ansatz solves 
the equations of motion for $\sigma_2$ and $\sigma_3$ and moreover because $(\partial\sigma_2)^2=(\partial\sigma_3)^2$
it is consistent to have $ \lambda_2=\lambda_3=\gamma$. The remaining equations of motion lead to 
a first order ODE for $U$ and two second order ODEs for
$V$ and $\gamma$ given by:
\begin{align}\label{eqom}
U'&=\frac{1}{2V'}\left(2(1+2\cosh\gamma)-e^{-2V}k^2 \sinh^2\gamma+U(\gamma'^2-2V'^2)\right)\,,\nn
V''&=-V'^2-\frac{1}{2}\gamma'^2\,,\nn
U\gamma''&=(-2+e^{-2V}k^2\cosh\gamma)\sinh\gamma-(U'\gamma'+2UV')\gamma'\,.
\end{align}

As $r\to \infty$ we demand that the solutions approach the $AdS_4$ solution with the following asymptotic behaviour 
    \begin{align}\label{bcs}
U=r^2+...\,,\qquad
e^{2V}=r^2+\dots\,,\qquad
\gamma=\frac{\Gamma}{r}+\dots\,.
\end{align}
It will be convenient to refer to $\Gamma$ and $k$ as `deformation parameters' in the following.
For fixed dimensionless parameter $\Gamma/k$, by solving the ODEs with prescribed boundary conditions in the IR,
we can then obtain the sub-leading terms in the expansion \eqref{bcs} and these can be used to parametrise 
`expectation values' of the dual operators. Viewing \eqref{eq:tLag2} from a bottom-up context this is the appropriate language 
to describe the RG flow when $\gamma$ is dual to an operator with scaling dimension $\Delta=2$. 
However, in the top-down context it is important to note that the ansatz for the complex scalars is written in terms of the $X_i$ and $Y_i$ as
\begin{align}\label{xandy}
X_2&= \gamma(r)\cos(kx),\qquad Y_2= \gamma(r)\sin(kx)\,,\nn
X_3&= \gamma(r)\cos(ky),\qquad Y_3= \gamma(r)\sin(ky)\,,
\end{align}
with $X_1=Y_1=0$.
Now supersymmetry implies that the scalars $X_i$ and the pseudoscalars $Y_i$ 
are associated with operators of scaling dimension $\Delta=1$ and $\Delta=2$, respectively. 
The parameter $\Gamma$ therefore describes deformations of 
two pseudoscalar operators with spatial dependence given by $\sin k x$ and $\sin k y$. However, the deformations of
two scalar operators, which have spatial dependence $\cos kx$ and $\cos ky$, are given by the sub-leading terms in \eqref{bcs}.
It is precisely this tuning of the deformation parameters of these operators that allows us to construct the RG flows by solving
a system of ODE's. It would be interesting to extend our solutions away from this tuned situation, but that will necessarily involve
solving partial differential equations and this will be left for future work. We also return to this issue below when
we discuss finite temperature solutions.

The solutions that we construct are in the $U(1)^4$ invariant bosonic sector of $N=8$ gauged supergravity. As such, after being uplifted on
the $S^7$ to obtain solutions of $D=11$ supergravity, they will survive the cyclic quotient of the $S^7$ and hence are relevant for the $N=6$ ABJM theory \cite{Aharony:2008ug}. The scalar fields $X_i$ are dual to operators that are scalar bilinears, while the pseudoscalar fields $Y_i$ are dual
to operators that are fermion bilinears. More precisely, under the $SU(4)\times U(1) \subset SO(8)$ holographically identified with the global symmetries of the ABJM theory, the $X_i$ and $Y_i$  each transform in a $ {\bf 15_0}  $ representation and are thus dual to operators schematically of the form
\begin{align}
\mathcal{O}_{\phi\phi} \sim \mathrm{Tr}\left( \phi^\dagger_{\bar A}\,{\phi}^B - \frac{1}{4}\delta_{\bar A}^B \,\phi^\dagger{\phi}\right)\,,\qquad
\mathcal{O}_{\psi\psi} \sim \mathrm{Tr}\left( \psi^\dagger_{\bar A}\,{\psi}^B - \frac{1}{4}\delta_{\bar A}^B \,\psi^\dagger{\psi}\right)\,,
\end{align}
respectively, where $\phi$,$\psi$ are scalar and fermion fields of the ABJM theory, respectively.
The RG flows are being driven by spatially modulated deformations of these operators and \eqref{xandy} shows that
this breaks spatial isotropy in the $(x,y)$ directions.

\subsection{Perturbative deformations}\label{pertcon}
The RG flow solutions that we have constructed depend on the dimensionless deformation parameter $\Gamma/k$.
For small deformations, $\Gamma/k\ll 1$, we obtain some important insight by solving the equations as a perturbative expansion
about the $AdS_4$ vacuum. 
At leading order in $\Gamma/k$ we can easily solve the linearised equation of motion for $\gamma$. 
Choosing the integration constants so that the solution is both regular at the Poincar\'e horizon and with boundary conditions as in \eqref{bcs} we find
\begin{align}
\gamma(r)=\frac{k}{r}{e^{-k/r}}\left({\Gamma}/{k}\right)+\mathcal{O}\left({\Gamma}/{k}\right)^2\,.
\end{align}
This solution will back react on the metric at order $(\Gamma/k)^2$ and explicit expressions can be obtained 
subject to the appropriate boundary conditions. We find 
\begin{align}\label{eq:pertsol}
U&=r^2[1+\left(\frac{k}{4r}e^{-2k/r}\right)\left({\Gamma}/{k}\right)^2+\mathcal{O}\left({\Gamma}/{k}\right)^3]\,,\nn
%e^{2V}&=r^2[1+\frac{1}{16}\left(2-e^{-2k/r}(2+\frac{4k^2}{r^2})\right)\left({\Gamma}/{k}\right)^2+\mathcal{O}\left({\Gamma}/{k}\right)^3]\,.\nn
e^{2V}&=r^2[1+\frac{1}{8}\left(1-e^{-2k/r}(1+\frac{2k^2}{r^2})\right)\left({\Gamma}/{k}\right)^2+\mathcal{O}\left({\Gamma}/{k}\right)^3]\,.
\end{align}

In the far IR, as $r\to 0$, the metric rapidly approaches the same $AdS_4$ solution that appears in the UV, 
with the scale of the approach set by $k$.
The only difference between the $AdS_4$  solutions in the UV and the IR is that there is a renormalisation of relative length scales.
In the UV we can define the ratio $\chi_{UV}=\lim_{r\to \infty}U^{1/2}/e^V$ 
%as $r\to \infty$
and similarly
$\chi_{IR}=\lim_{r\to 0}U^{1/2}/e^V$ in the IR, both of which are invariant under scalings of the radial coordinate.
%as $r\to 0$
Following \cite{Gubser:2009gp} we can then define the RG flow invariant, $n$, (sometimes called 
the `index of refraction' for the RG flow) as
$n\equiv \frac{\chi_{UV}}{\chi_{IR}}$. Since in the parametrisation we are using $U\to r^2$ both as $r\to\infty$ and $r\to 0$ we have
\begin{align}\label{refind}
n%\equiv \frac{c_{UV}}{c_{IR}}
=\frac{e^{V}(r\to 0)}{e^{V}(r\to \infty)}
=1+\frac{1}{16}\frac{\Gamma^2}{k^2}+\mathcal{O}\left({\Gamma}/{k}\right)^3\,.
\end{align}

The recovery of conformal invariance that we see\footnote{The perturbative argument we used above was also used to argue for boomerang RG flows in the context of other examples of CFT deformations which break translations \cite{Chesler:2013qla,Donos:2014gya,Donos:2016zpf}. One context it does not apply is if the linearised deformation gives rise at higher orders in the perturbative expansion
to additional sources with non-vanishing zero modes (i.e. the integral of the source over a spatial period is non-vanishing). It  also does not apply to the type IIB linear dilaton and linear axion solutions of \cite{Jain:2014vka} and \cite{Azeyanagi:2009pr}, respectively.}
for small values of $\Gamma/k$ 
is associated with the fact that the operators 
used in the deformation have vanishing spectral weight at low energies. To determine what happens
for larger values of
$\Gamma/k$ it is necessary to solve the equations of motion numerically. Our numerical results, summarised below, 
indicate that for arbitrarily large values of $\Gamma/k$ the RG flows are all boomerang flows.
In addition, we find that for large enough $\Gamma/k$
all of the RG flow solutions successively approach,
for intermediate values of the radial coordinate $r/k$, 
two intermediate scaling behaviours before hitting the $AdS_4$ behaviour in the far IR. 

\section{Intermediate Scaling Solutions}\label{intscal}
The two intermediate scaling behaviours that we observe in the boomerang RG flows are, somewhat surprisingly, 
not associated with exact hyperscaling solutions of the equations of motion coming from the Lagrangian $\mathcal{L}$ in  
\eqref{eq:tLag2}. In this section we explain their orgin.

\subsection{First intermediate scaling regime: the $k=0$ flow}\label{keqzf}
The first intermediate scaling regime that we observe on the way to the IR is governed by large values of the field $\gamma$ and, moreover, is such that the terms involving $k$ play a sub-dominant role in the equations of motion. Since the breaking of translation invariance is sub-dominant
the first intermediate scaling behaviour is approximately Lorentz invariant.

Let us therefore consider Lorentz invariant RG flow solutions of the equations of motion \eqref{eqom} with $k=0$ (i.e. $\sigma_2=\sigma_3=0$) 
and $e^{2V}=U$. 
We look for solutions that approach $AdS_4$ in the UV with expansion \eqref{bcs}. 
While there aren't any exact hyperscaling violation solutions to the equations of motion that we can map onto in the IR, we have numerically constructed
RG flow solutions
that approach the following singular behaviour\footnote{The subleading corrections are more easily obtained by switching radial coordinate so that the metric is of the form $ds^2=dR^2+e^{2W}(-dt^2+dx^2+dy^2)$ and we then find that as $R\to 0$ we have
$e^{2W}=R^4[1+\frac{2}{21}R^2+o(R^4)]$ and $e^\gamma=\frac{10}{R^2}[1+\frac{1}{42}R^2+o(R^4)]$.}
 in the far IR as $r\to 0$:
\begin{align}\label{fisr}
U=e^{2V}=L_I^{-2}r^{4/3}+\dots,\qquad
e^\gamma=e^{\gamma_0}r^{-2/3}+\dots\,,
\end{align}
where $L^2_I=(10/9)e^{-\gamma_0}$.
Notice that the field $\gamma$ is diverging as $r\to 0$.
In particular, the IR behaviour of the $k=0$ flow
is approaching that of solutions with hyperscaling violation, with a running scalar, 
similar to the flows in \cite{Gubser:2000nd} 
(see also \cite{Chamblin:1999ya}).
To see this more explicitly, we introduce a new radial coordinate $\rho=(1/3L_I)r^{-1/3}$ and, after suitably 
scaling $t,x,y$, we find that we can write the leading form of the IR metric, now located at $\rho\to\infty$
as:
\begin{align}\label{genmet}
ds^2=\rho^{-(2-\theta)}\left(-\rho^{-2(z-1)}d\bar t^2+d\rho^2+d\bar x^2+d\bar y^2\right)\,,
\end{align}
with dynamical exponent $z=1$, associated with Lorentz invariance, and hyperscaling violation exponent $\theta=-2$ (in the parametrisation of \cite{Huijse:2011ef}). Note that under the scaling
\begin{align}
t\to \mu^z t,\quad (x,y)\to \mu (x,y),\quad \rho\to \mu \rho\,,
\end{align}
the general metric \eqref{genmet} scales as $ds^2\to \mu^\theta ds$. Furthermore, if one heats up these solutions
one finds that the entropy density scales like $s\propto T^{(2-\theta)/z}=T^{4}$, a scaling we will see in the finite temperature solutions
that we discuss in section \ref{finitetemp}. Finally, we note that
we have checked that the $k=0$ flow does not preserve supersymmetry.

\subsection{Second intermediate scaling regime}
The second intermediate scaling regime that arises in the RG flows is governed by large values of the field $\gamma$ but now
the terms involving $k$ play a comparable role in the equations of motion. Thus, in contrast to the first intermediate scaling regime,
this scaling behaviour breaks translation invariance.

Perhaps the simplest way to describe this behaviour is to consider the following 
auxiliary Lagrangian, $\hat{\mathcal{L}}$, which approximately governs the behaviour of solutions in regions of spacetime where
the field $\gamma$ is getting large:
%\begin{equation}\label{eq:tLag2a}
 %\hat{\mathcal{L}} = \sqrt{-g}\Big(R-(\partial\gamma)^2-\frac{e^{2\gamma}}{8} \left[ (\partial\sigma_2)^2+(\partial\sigma_3)^2\right]
 %+8g^2e^\gamma\Big)\,.
%\end{equation}
\begin{equation}\label{eq:tLag2a}
\mathcal{L} = R-\frac{1}{2}(\partial\lambda_2)^2-\frac{1}{8}e^{\lambda_2}(\partial\sigma_2)^2-\frac{1}{2}(\partial\lambda_3)^2-\frac{1}{8}e^{\lambda_3}(\partial\sigma_3)^2+(e^{\lambda_2}+e^{\lambda_3})\,.
\end{equation}
Within the ansatz \eqref{ansatz}, there exists an exact hyperscaling violation solution for this auxiliary theory, which was first given in \cite{Donos:2014uba} (see also \cite{Gouteraux:2014hca}). It takes the form, for all values of $r$,
\begin{align}\label{sndis}
U=L_{II}^{-2}r^{8/3},\qquad e^{2V}=e^{2v_0}r^{2/3},\qquad
e^\gamma=e^{\gamma_0}r^{2/3}\,,
\end{align}
where $L_{II}^2=(28/9)e^{-\gamma_0}$ and $e^{\gamma_0}=6e^{2v_0}/{k^2}$.
By introducing a new radial coordinate $\rho=(9L_{II}^2/4)r^{-2/3}$ and suitably scaling $t,x,y$, we can write the metric in
the form of \eqref{genmet}
with dynamical exponent $z=5/2$ and hyperscaling violation exponent $\theta=1$. In the second
intermediate scaling regime the RG flow solutions approach the behaviour as in \eqref{sndis} with large values of $\gamma$ (i.e. with $r$ in 
\eqref{sndis} going to $\infty$.) If one heats up these solutions
one finds that the entropy density scales like $s\propto T^{2/5}$.

\section{The RG flows}\label{rgflows}
We now summarise the RG flows that we have constructed numerically. They are solutions within the ansatz
\eqref{ansatz} and solve the equations of motion \eqref{eqom}. 
In the far IR, as $r\to 0$, they approach $AdS_4$ with expansion given by
\begin{align}\label{irbcs}
U&=r^2\left(1+e^{-2\frac{k}{r{c_V}}}\left(\frac{{c_{V}}c^{2}_{\gamma}}{4 k r}\,\right) +\cdots \right)\,,\nn
e^{2V}&=r^{2}c_{V}^2\left(1- e^{-2\frac{k}{r{c_V}}} \left( \frac{1}{4 r^{2}}+\frac{ c_{V}^2}{8k^{2}}\right)c_{\gamma}^{2}+\cdots\right)\,,\nn
\gamma&=\frac{c_{\gamma}}{r}e^{-\frac{k}{r{c_V}}}+\cdots\,,
\end{align}
depending on two integration constants, $c_\gamma$, $c_V$.
In the UV, as $r\to\infty$, we assume that the solutions approach $AdS_4$ with the following expansion
\begin{align}\label{uvfullexp}
U&=(r+r_{+})^{2}\,\left(1-\frac{1}{2}\frac{\Gamma^{2}}{(r+r_{+})^{2}}+\frac{M}{(r+r_{+})^{3}}+\cdots \right)\notag\\
e^{2V}&=(r+r_{+})^{2}\,\left(1-\frac{1}{2}\frac{\Gamma^{2}}{(r+r_{+})^{2}}-\frac{2}{3}\frac{\Gamma\, \hat{\Gamma}}{(r+r_{+})^{3}}+\cdots \right)\notag\\
\gamma&=\frac{\Gamma}{r+r_{+}}+\frac{\hat\Gamma}{(r+r_{+})^{2}}+\dots\,,,
\end{align}
with the appearance of $r_+$ related to the fact that we have set the IR at $r=0$.
Our boundary conditions will be to hold fixed the dimensionless ratio $\Gamma/k$. The three constants of integration $\hat{\Gamma}$, $r_{+}$ and $M$ appearing in \eqref{uvfullexp} will be fixed by demanding regularity in the IR part of the geometry. For example, the perturbative solution \eqref{eq:pertsol}, which has a smooth IR limit at $r=0$, has $r_{+}=\Gamma^{2}/(8 k)$, $\hat{\Gamma}=-k\,\Gamma$ and $M=k\Gamma^{2}/2$. More generally, for the solutions that we will obtain numerically these three constants can be fixed by shooting both from the UV and the IR and then matching at intermediate values of the radial coordinate. The equations of motion \eqref{eqom} that we will be integrating will require five constants to be fixed in this way. Therefore, two of them will have to come from the IR expansion. For the RG flows at $T=0$ these two constants are
$c_{\gamma}$ and $c_{V}$ in \eqref{irbcs}. For the finite temperature solutions, which we discuss
in subsection \ref{finitetemp}, the two extra constants will come from an analytic expansion around a regular horizon
which will be located at the fixed position $r=0$, 
(again associated with the appearance of $r_+$ in \eqref{uvfullexp}),
which is convenient for the numerics.

For small $\Gamma/k$ the solutions are well approximated by the perturbative solutions that we constructed in 
section \ref{pertcon}. After sufficiently increasing the value of $\Gamma/k$ we then start to approach the first intermediate scaling regime, governed
by the IR behaviour of the $k=0$ flow \eqref{fisr}, thus approximately recovering Lorentz invariance. 
This first intermediate scaling regime arises because, since we are deforming by a relevant operator, the dimensionless
deformation parameter, $\Gamma/k$, involves $k$. In particular, one can expect that the $\Gamma/k\to\infty$ behaviour should approach
the $k=0$ with $\Gamma\ne 0$.

\begin{figure}[t!]
\includegraphics[width=0.5\textwidth]{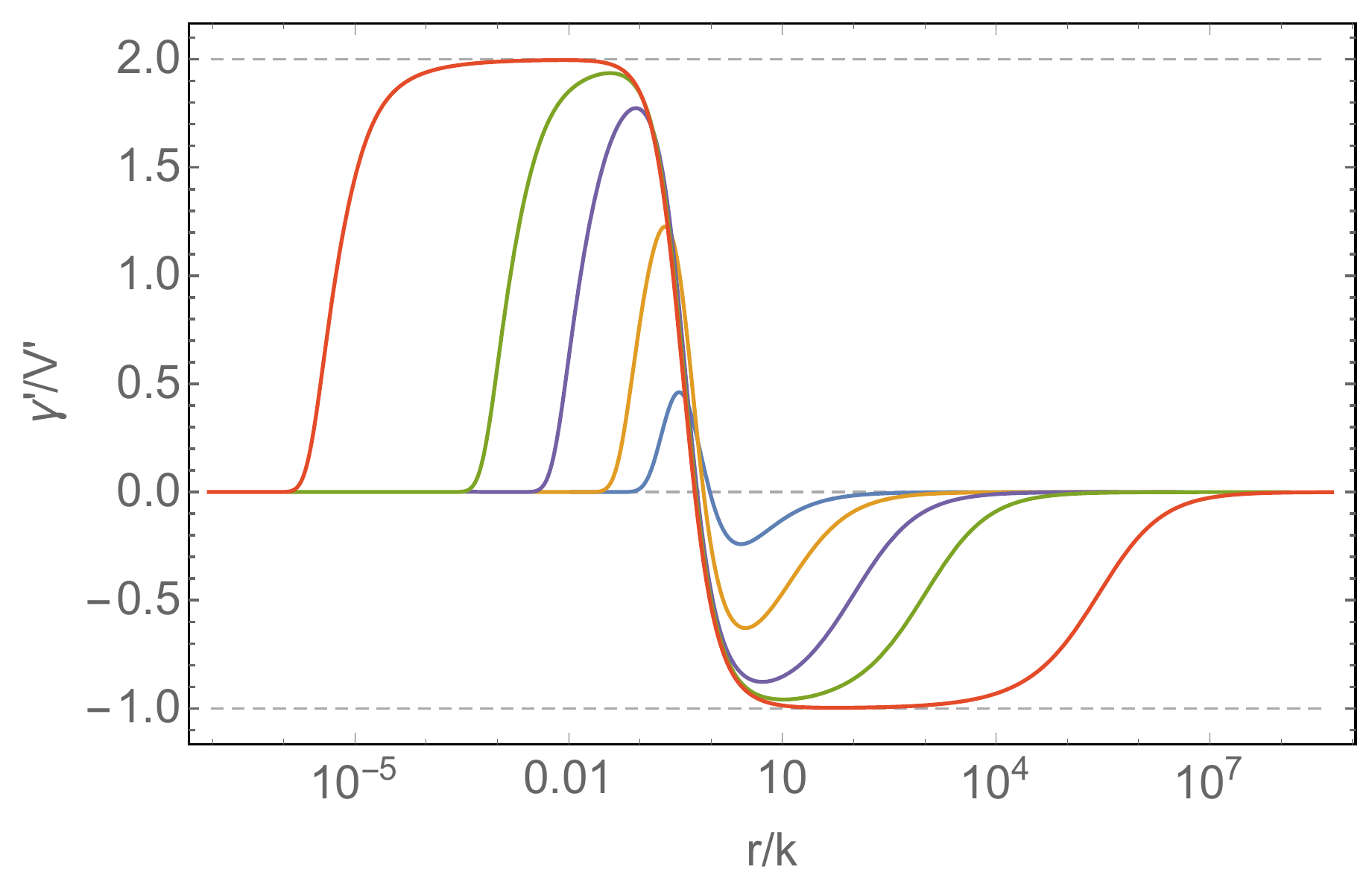}%\,,\quad
\includegraphics[width=0.5\textwidth]{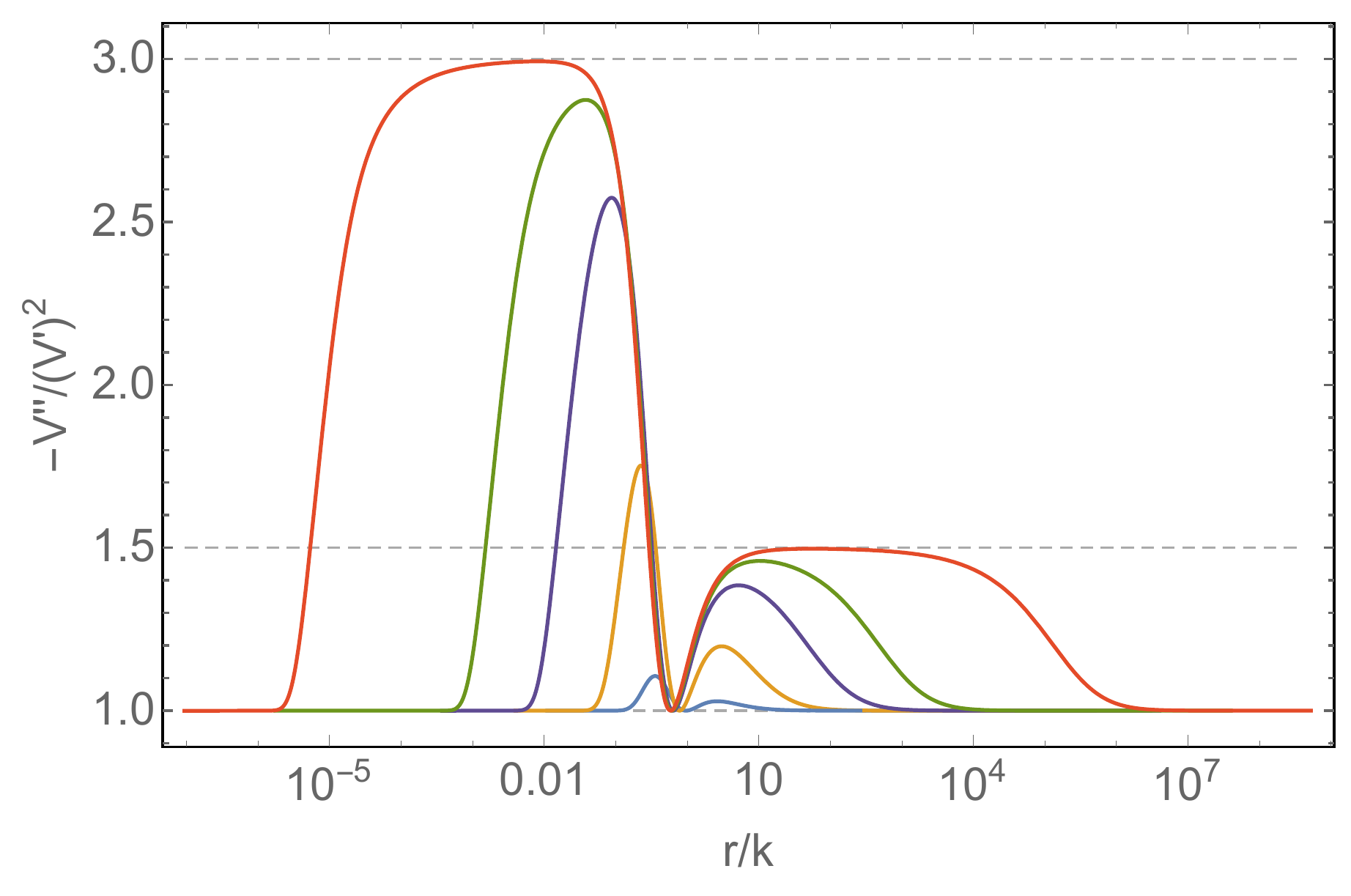}
\center{\includegraphics[width=0.5\textwidth]{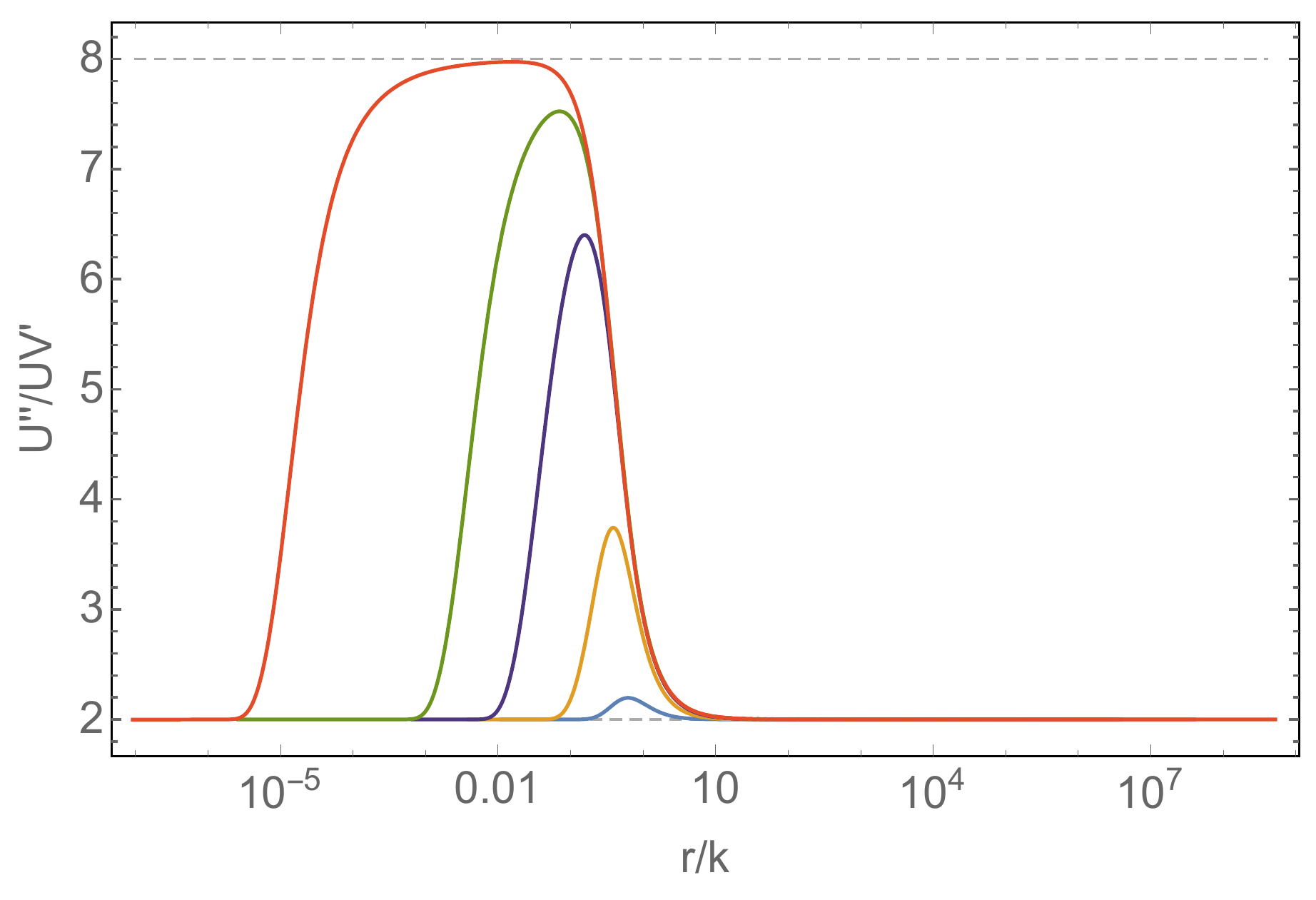}}
\caption{Plots of various functions associated with the RG flows, as functions of the dimensionless radial coordinate $r/k$, for various values 
of the dimensionless deformation parameter $\Gamma/k$: blue ($\Gamma/k=1$), orange ($\Gamma/k=10$), purple  ($\Gamma/k=10^2$), green ($\Gamma/k=10^3$) and red
($\Gamma/k=2.6\times10^5$). The plots demonstrate the boomerang RG flow from $AdS_4$ in the UV to $AdS_4$ in the IR for all values of $\Gamma/k$. 
For sufficiently large values of  $\Gamma/k$, on the way to the IR the flows approach two intermediate scaling regimes.
\label{fig2}}
\end{figure}

Interestingly, for the same large values of $\Gamma/k$, as we go further into the IR, we
also approach the second intermediate scaling region \eqref{sndis} with $z=5/2$. 
A convenient way\footnote{The scaling behaviour displayed by these functions is invariant under shifts of the radial coordinate by a constant.}
 of displaying the scaling behaviour is to plot 
$\gamma'/V'$, $-V''/(V')^2$ and $U'/(UV')$ as functions of the dimensionless radial coordinate $r/k$.  
In particular, for the first intermediate scaling regime \eqref{fisr} these functions should approach 
$-1$, $3/2$ and $2$, respectively. Similarly for the second intermediate scaling regime \eqref{fisr} these functions should approach 
$2$, $3$ and $8$, respectively. Furthermore, for boomerang RG flows they should approach $0$, $1$ and $2$, respectively, both in
the UV and in the IR. In figure \ref{fig2} we demonstrate the behaviour of these functions for 
several representative values of $\Gamma/k$ and we clearly see the boomerang RG flows and the appearance of the intermediate scaling regimes. 

\begin{figure}[t!]
\centering
\includegraphics[width=0.5\textwidth]{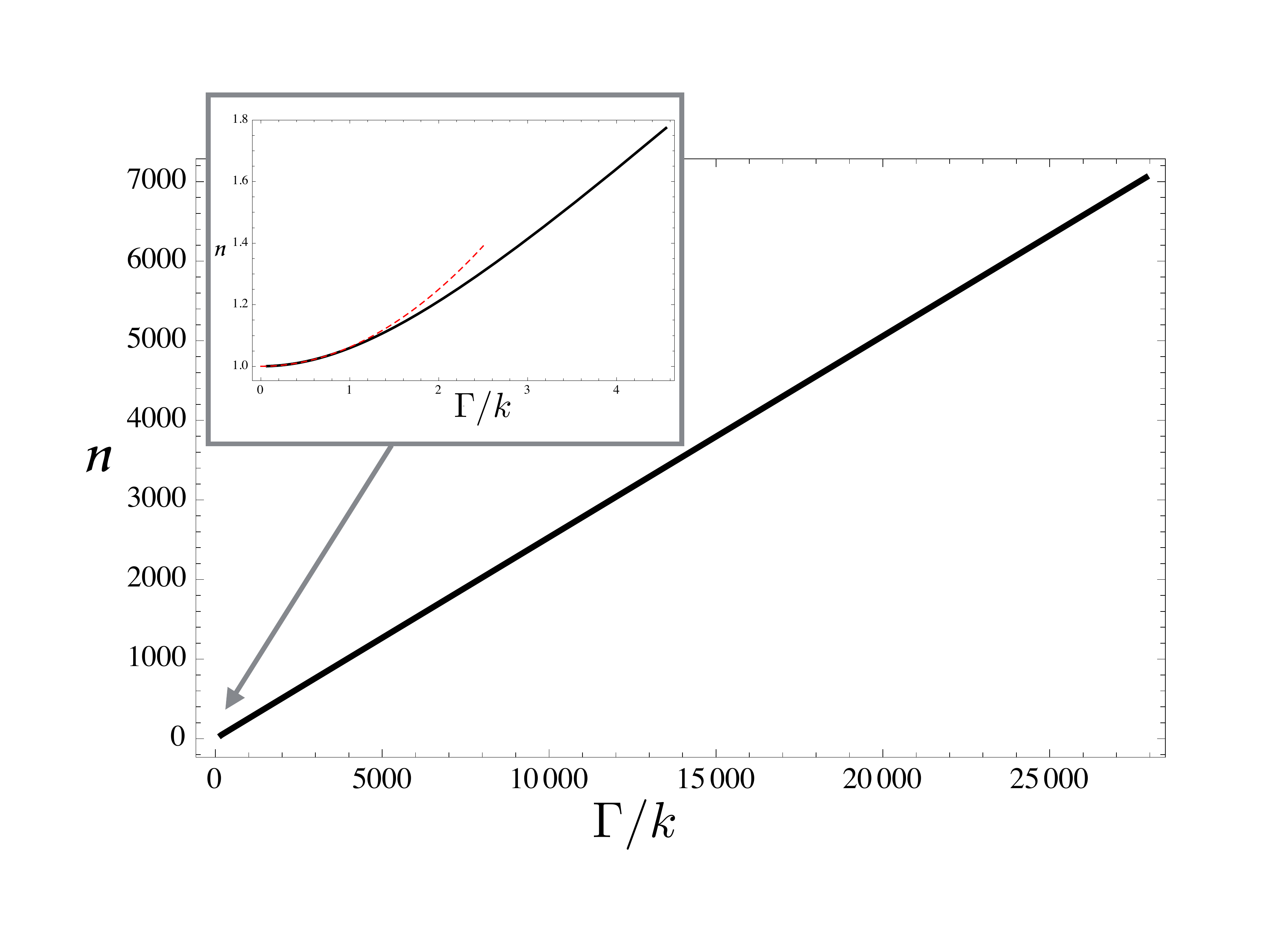}
\caption{Plot of the RG flow invariant $n$, defined in \eqref{refind}, versus deformation parameter $\Gamma/k$, for the boomerang RG flows.
The inset shows excellent agreement with the perturbative result given in \eqref{refind} for $\Gamma/k\ll 1$.
\label{fig3}}
\end{figure}

By scanning over different values of $\Gamma/k$ we can also determine the behaviour of the RG flow invariant $n$, as defined in \eqref{refind}, and our results are presented in figure \ref{fig3}. For small
values of $\Gamma/k$ we see that $n-1$ depends quadratically on 
$\Gamma/k$, as expected from the perturbative analysis. For large values $\Gamma/k$ we find that $n$ asymptotes to a linear dependence
of the form $n\sim 0.253(\Gamma/k)$.
 
\subsection{Finite temperature}\label{finitetemp}

The intermediate scaling regimes that we have found in the RG flows, for large enough values of $\Gamma/k$, should also manifest themselves at non-zero temperature $T$, for $T/k\ll\Gamma/k$. We have constructed finite temperature black hole solutions
by changing the IR boundary conditions from $AdS_4$, as in \eqref{irbcs}, to a regular black hole Killing horizon located at $r=0$, with
\begin{align}\label{irbcsbh}
U&=4\pi T\,r+\frac{1}{2}e^{-2 V_{1+}} k^2\sinh^2\gamma_{+}\,r^2+\cdots \,,\nn
V&=V_{1+}+\frac{1}{8\pi T}(2+4\cosh\gamma_+ - e^{-2 V_{1+}} k^2\sinh^2\gamma_+)\,r+\cdots\,,\nn
\gamma&=\gamma_++\frac{1}{8\pi T}(4\sinh\gamma_+-e^{-2 V_{1+}}k^2\sinh 2 \gamma_+)\,r+\cdots \, ,
\end{align}
where $T$ is the temperature which we will be holding fixed.
%\begin{align}\label{irbcsbh}
%U&=4\pi T(r-r_+)+\frac{1}{2}(r-r_+)^2 e^{-2 V_{1+}} k^2\sinh^2\gamma_{+}... \,,\nn
%V&=V_{1+}+V_{2+}(r-r_+)+...\,,\nn
%\gamma&=\gamma_++\frac{V_{2+}(4\sinh\gamma_+-e^{-2 V_{1+}}k^2\sinh 2 \gamma_+) }{2 +4\cosh\gamma_+ -e^{-2 V_{1+}} k^2\sinh^2\gamma_+}(r-r_+)+... \, ,
%\end{align}
%where the temperature is given by
%\begin{equation}\label{temp}
%T = \frac{2+4\cosh\gamma_+ - e^{-2 V_{1+}} k^2\sinh^2\gamma_+}{8\pi V_{2+}}.
%\end{equation}
The two constants of integration $\gamma_{+}$ and $V_{1+}$ are used to find a unique solution of the equations of motion \eqref{eqom} upon matching with the three constants that we discussed below \eqref{uvfullexp}.

\begin{figure}[t!]
\centering
\includegraphics[width=0.5\textwidth]{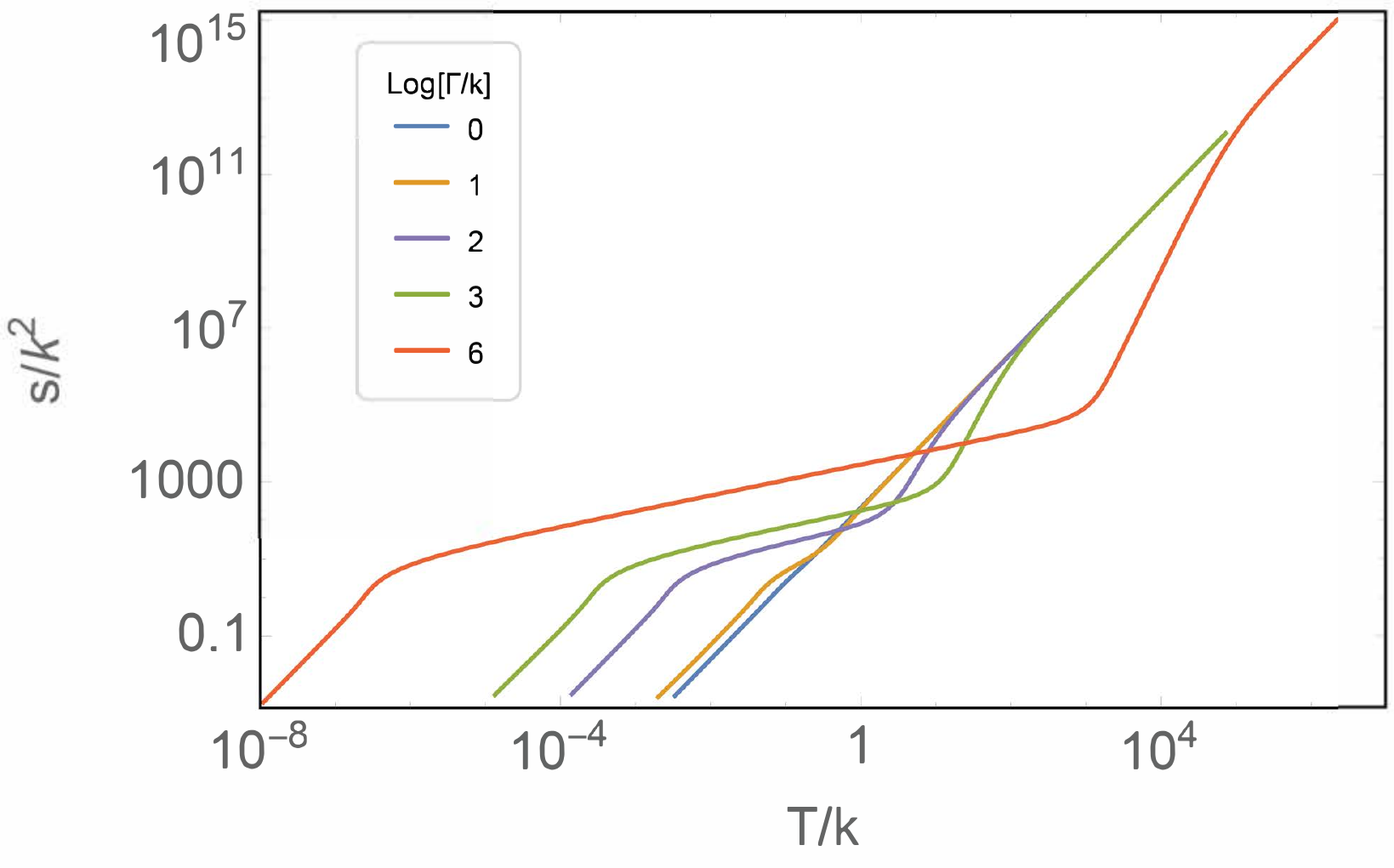}%\quad
\includegraphics[width=0.5\textwidth]{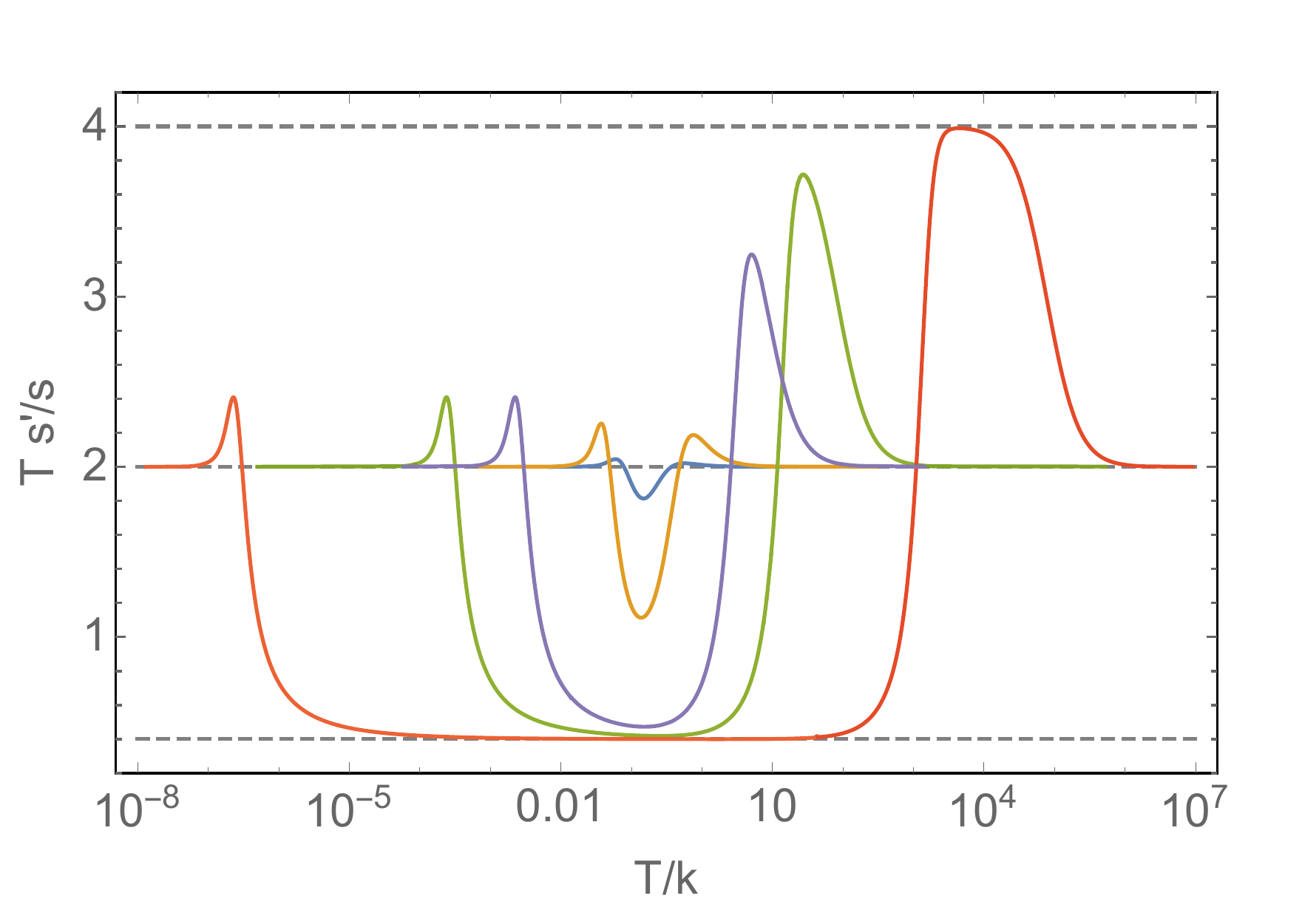}
\caption{Behaviour of the entropy density, $s$, as a function of temperature,
%$s/k^2$ as  $Ts'/s$, where $s$ is the entropy density and $T$ is the temperature, versus $T/k$ 
for various boomerang RG flows: blue 
($\Gamma/k=1$), orange ($\Gamma/k=10$), purple ($\Gamma/k=10^2$), green ($\Gamma/k=10^3$) and red
($\Gamma/k=10^6$). In the UV and in the IR we have $Ts'/s\to 2$, associated with $AdS_4$. We also
see, for large enough values of $\Gamma/k$, the appearance of two intermediate scaling regimes that are
governed by the hyperscaling violation solutions
which have $Ts'/s=4$ and $Ts'/s=2/5$.
\label{fig4}}
\end{figure}

For temperatures $T/k\ll\Gamma/k$ we can anticipate that there are 
intermediate regimes of low temperature where the solutions approach that of a hyperscaling
violation black hole with $z=5/2$, $\theta=1$ and then, for large temperatures, a hyperscaling
violation black hole with $z=1$, $\theta=-2$. Correspondingly, this should give rise to an associated scaling of
thermodynamic quantities. For example, the scaling of the entropy density
should begin as $s\sim T^2$ for low temperatures, associated with the $AdS_4$ IR behaviour.
Then, as we increase the temperature we should successively see $s\sim T^{2/5}$ followed by $s\sim T^4$,
corresponding to the two hyperscaling regimes, and finally end up with $s\sim T^2$ for very high temperatures
corresponding to the $AdS_4$ region in the UV. These features are clearly displayed for a range of $\Gamma/k$ as shown in figure
\ref{fig4}.

We would like to highlight an important subtlety concerning these finite temperature black hole solutions. By construction
we find a one parameter family of black hole solutions, labelled by $T/k$, while holding
the parameter $\Gamma/k$ fixed. As we discussed below \eqref{xandy}, this means that we are holding fixed
the deformation parameters for the two pseudoscalar operators, dual to the $Y_i$, as well as 
the expectation value of the two scalar operators, dual to the $X_i$.
It is precisely for this particular mixed thermodynamic ensemble that we are able to construct black hole solutions by solving
ordinary differential equations.

\subsection{Spectral weight of operators in the RG flows}
We now return back to the RG flows at zero temperature and
analyse the behaviour of some correlation functions involving scalar operators.
In particular, we will show how
the intermediate scaling regimes 
can also lead to scaling behaviour appearing in various spectral functions of the dual field theory,
for certain ranges of
intermediate frequencies. We will also see that there can be an interesting kind of universality in which operators of different scaling dimensions in the UV exhibit the same scaling at intermediate scales.
Some additional interesting features will be highlighted as we proceed. 

In general, given that spatial translations have been explicitly broken, we need to
consider linearised perturbations about the RG flows that involve solving partial differential equations.
However, there are some correlation functions that can be obtained by solving ordinary differential equations, and this is
what we will study here.
Specifically,
we start by considering a bulk scalar field $\phi$ whose linearised equation of motion in the background geometry is
given by the Klein-Gordon equation
\begin{align}\label{kg}
(\nabla^2-m^2)\phi=0\,.
\end{align}
A specific case that we will focus on is when $m^2=-2$: this arises in the STU model, given in \eqref{eq:Lag},
for the scalar field $\lambda_1$ with $\sigma_1=0$ (i.e. $X_1$ in \eqref{xandyo}), which is dual to an operator with $\Delta=1$
and also $\lambda_1$ with $\sigma_1=\pi/2$ (i.e. $Y_1$ in \eqref{xandyo}), which is dual to an operator with $\Delta=2$.
As usual, the retarded Green's function $G^R(\omega)$ can be obtained by 
writing $\phi=e^{-i\omega t}\,\tilde{\psi}(r)$ and then solving \eqref{kg} with ingoing boundary conditions in the $AdS_4$ geometry in the far IR. This gives a radial equation for $\tilde\psi$ and the ratio of the
normalisable to the non-normalisable solutions in the UV, then gives ${G}^R(\omega)$. For example, for the $\Delta=1$ operator we
can expand at $r\to\infty$ as 
$\tilde{\psi}(r)=\psi_1(\omega)/r+\psi_2(\omega)/r^2+...$ and we have ${G}^R(\omega)\propto\psi_1(\omega)/\psi_2(\omega)$, while for the
$\Delta=2$ operator we have the same expansion with ${G}^R(\omega)\propto \psi_2(\omega)/\psi_1(\omega)$.

It is convenient to introduce a new radial coordinate, $z$, defined by
\begin{align}\label{zeddef}
z=-\int_{r}^{+\infty}\frac{dy}{U(y)}\,,
\end{align}
and we note that the UV is located at $z=0$ and the IR at $z=-\infty$. We then have that $v=t+z$ is the ingoing coordinate in Eddington-Finkelstein coordinate. Next, by writing $\tilde{\psi}=e^{-V}\,\psi$, we then deduce that the radial equation can be written in the Schr\"odinger form
\begin{align}\label{radeq}
-\partial_{z}^{2}\psi+\left(\mathcal{V}-\omega^{2}\right)\psi&=0\,,
\end{align}
where we have defined the effective potential
\begin{align}\label{eq:sch_potential}
\mathcal{V}=U\left( m^{2}+e^{-V}\partial_{r}\left( U\partial_{r}e^{V}\right)\right)\,.
\end{align}

Now for standard RG flows, which flow from the UV to another geometry with scaling behaviour in the far IR, matching arguments have been developed in
\cite{Donos:2012js}, generalising earlier work, including \cite{Faulkner:2011tm}, which show that for small frequencies the spectral function
$\text{Im}
{G}^R(\omega)$ is determined by the spectral function associated with the retarded Green's function\footnote{As explained in \cite{Donos:2012js}, in general
it is given by a sum of terms associated with various fields in the IR.} for the IR geometry, $\text{Im}\mathcal{G}^R(\omega)$.

We would like to know when something similar occurs for a background geometry with an intermediate scaling regime for $z_{1}<z<z_{2}$.
Specifically we want to determine when $\text{Im}{G}^R(\omega)$ exhibits scaling behaviour, for certain intermediate values of $\omega$, that is
fixed by spectral functions $\text{Im}\mathcal{G}^R_i(\omega)$, $i=1,2$, associated with one of the two intermediate scaling regimes. In order for this to occur we need to ensure that in the region
$z_{1}<z<z_{2}$, the solution of the radial equation \eqref{radeq}, which has ingoing boundary conditions imposed in the far IR at $z\to-\infty$,
is predominantly a solution in the intermediate scaling regime with ingoing boundary conditions imposed at $z=z_1$. In general this will
not be the case\footnote{For the solutions with intermediate scaling constructed in \cite{Bhattacharya:2014dea} it was numerically
shown that the conductivity exhibited intermediate scaling. Some matching arguments were
also discussed to explain this behaviour, but the sufficient conditions on the potential for when intermediate scaling appears,
that we identify here, were not discussed.} and the solution will also contain a significant admixture of a solution with outgoing 
boundary conditions imposed at $z=z_1$.

To proceed, for the scaling region $z_{1}<z<z_{2}$ we assume
\begin{align}\label{rangesc}
\left| \mathcal{V}(z_{1})\right| \ll \omega^{2} \ll \left| \mathcal{V}(z_{2})\right|\,,
\end{align}
and also demand that the potential satisfies
\begin{align}\label{potcond}
\left| \mathcal{V}(z)\right| <\left| \mathcal{V}(z_{1})\right| \ll \omega^{2}\,,\qquad \text{for}\qquad z<z_{1}\,.
\end{align} 
To see that this is sufficient to have intermediate scaling of the spectral function, we next split the $z$ interval into three regions:
\begin{align}
\psi^{(I)}(z)\approx\begin{cases}\psi^{(I)}(z),&\quad z<z_{1}\,, \\ \psi^{(II)},&\quad z_{1}<z<z_{2}\,, \\ \psi^{(III)},&\quad z>z_{2}\,. \end{cases}
\end{align}
In the region $z<z_{1}$, using the perturbative expansion parameter $\epsilon=\frac{\left|\mathcal{V}(z_{1})\right|}{\omega^{2}}$, we can develop
the following perturbative solution 
\begin{align}
\psi^{(I)}(z)= C_{in}(\omega)\,e^{-i \omega z}+C_{out}(\omega) \,e^{i \omega z}+\epsilon\,\psi^{(I)}(z)+\cdots\,,
\end{align}
where $C_{in}(\omega)$, $C_{out}(\omega)$ are constants. The infalling boundary conditions at $z=-\infty$ require that $C_{out}=0$. 
But this also shows that in the overlapping region around $z=z_{1}$, to leading order in $\epsilon$,  
we should impose approximate infalling boundary conditions on the matching solution $\psi^{(II)}(z)$. Thus, to leading order in $\epsilon$,
the solution $\psi^{(II)}(z)$ will be the usual perturbation in the scaling region $z_{1} < z < z_{2}$, with ingoing boundary conditions at
$z_1$. We can then invoke the matching arguments of \cite{Donos:2012js} to match onto the solutions in region $III$ and deduce that
for $\left| \mathcal{V}(z_{1})\right| \ll \omega^{2} \ll \left| \mathcal{V}(z_{2})\right| $, the spectral function $\text{Im}{G}^R(\omega)$
will be determined by $\text{Im}\mathcal{G}^R(\omega)$ where $\mathcal{G}^R(\omega)$ is the spectral function for the scaling solution in the region $z_{1}<z<z_{2}$. It is important to appreciate that in making this argument we do not need to know about the properties of $\mathcal{G}^R(\omega)$
for other values of $\omega$ and, for example, it is possible that it has instabilities which do not play a role.  
\begin{figure}[t]
\centering
\includegraphics[width=0.5\textwidth]{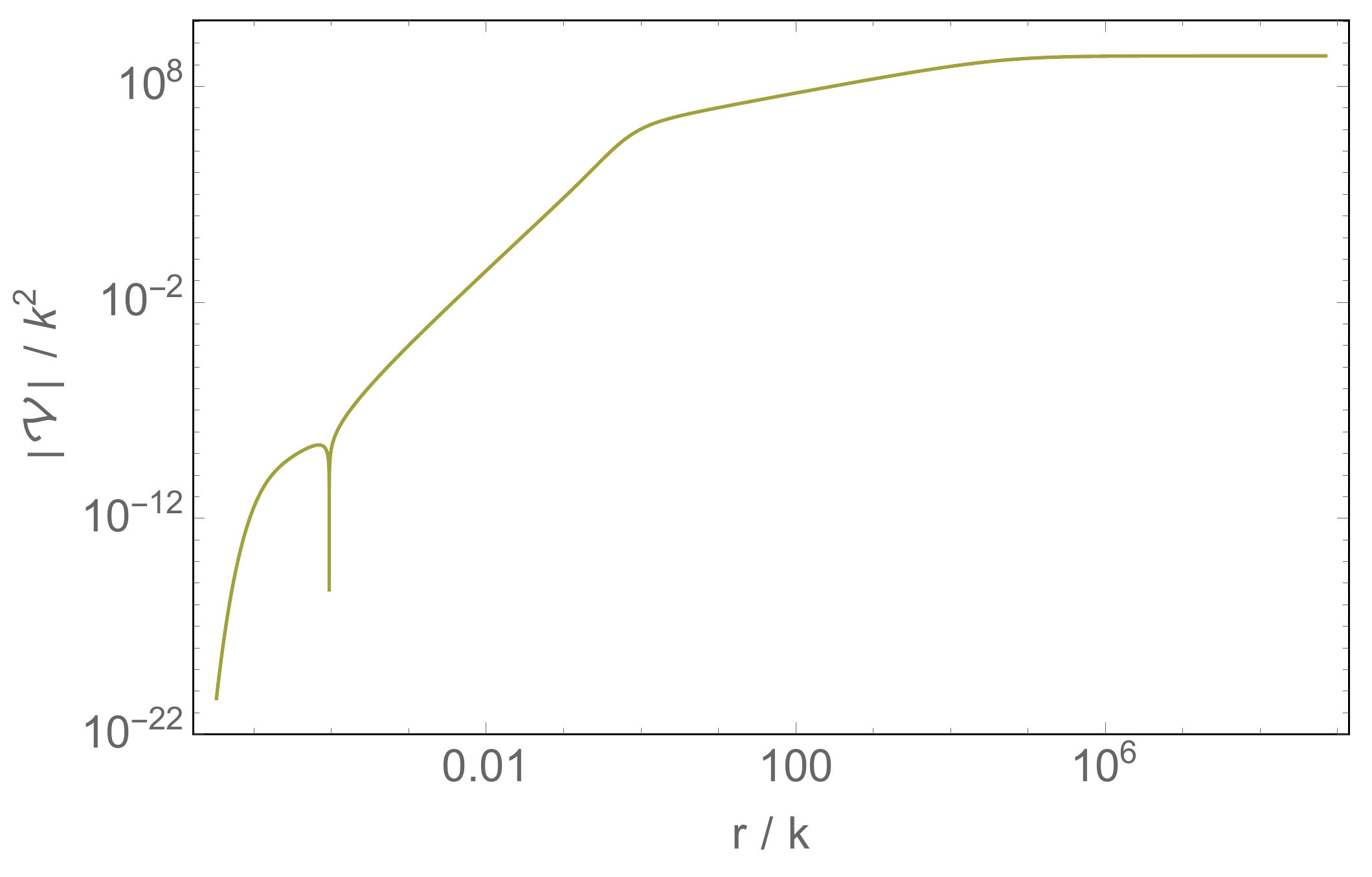}
\caption{Plot of the Schr\"odinger potential 
$|\mathcal{V}|$, defined in \eqref{eq:sch_potential}, for a massive scalar field with $m^2=-2$. 
Note that $\mathcal{V}$ has a zero at $r/k\sim 9.5\times 10^{-5}$.
The deformation parameter is
$\Gamma/k=0.7\times 10^5$ and the intermediate scaling behaviour is between $10^{-4}<r/k<1$ and $1<r/k<10^{4}$. The plot shows a clear separation of scales for the values of the potential in these regions.\label{fig:pot_scaling}}
\end{figure}

We can illustrate these ideas for the boomerang RG flows for the special cases mentioned above, with $m^2=-2$ and 
quantised so that $\Delta=1$ or $\Delta=2$. In figure \ref{fig:pot_scaling} we have plotted 
the Schr\"odinger potential $|\mathcal{V}|/k^2$ against $r/k$ for the RG flow with 
$\Gamma/k\sim 0.7\times 10^5$. The plot shows that the potential has a power law behaviour for the intermediate scaling regions with $1<r/k<10^{4}$
and $10^{-4}<r/k<1$. The plot also shows that the condition \eqref{potcond} is satisfied for both regions and hence we expect that there is
an intermediate scaling behaviour for the spectral function that is governed by the two hyperscaling violation solutions.

For $1<r/k<10^{4}$ we have $10^6 < |\mathcal{V}| / k^2 < 10^9$ and hence we
expect from \eqref{rangesc} that there will be intermediate scaling governed by the hyperscaling violation 
geometry \eqref{fisr} (associated with the $k=0$ flow) for the range of frequencies $10^6 \ll(\omega/k)^2 \ll10^9$. 
It is important to notice that in the intermediate scaling regime, the mass term in the Schr\"odinger potential \eqref{eq:sch_potential} 
is sub-dominant compared to the other term; this can easily be deduced by taking $r\to 0$ in \eqref{fisr}.
Taking this point into consideration, a straightforward calculation shows that for the hyperscaling violation geometry \eqref{fisr} we
have $\text{Im}\mathcal{G}^R_I(\omega)\sim \omega^{7/5}$ for small $\omega$ and hence in the boomerang flow
we expect to have the scaling $\text{Im}{G}^R(\omega)\sim\omega^{7/5}$ for $10^6 \ll (\omega/k)^2 \ll 10^9$.

Similarly, for $10^{-4}<r/k<1$ we have $10^{-8} <   |\mathcal{V}| / k^2 < 10^6$ and hence we expect intermediate scaling governed by the hyperscaling
violation geometry \eqref{sndis} for the range $10^{-8} \ll  (\omega/k)^2 \ll 10^6$.
Once again the mass term in the Schr\"odinger potential \eqref{eq:sch_potential} is sub-dominant compared to the other term in the intermediate scaling regime governed by the hyperscaling violation geometry \eqref{sndis}. Now for \eqref{sndis} a calculation shows that
$\text{Im}\mathcal{G}^R_{II}(\omega)\sim \omega^{5}$ for small $\omega$ and hence for the boomerang flow
we expect to have the scaling $\text{Im}{G}^R(\omega)\sim\omega^{5}$ 
for $10^{-8} \ll  (\omega/k)^2 \ll 10^6$.

\begin{figure}[t!]
\includegraphics[width=0.5\textwidth]{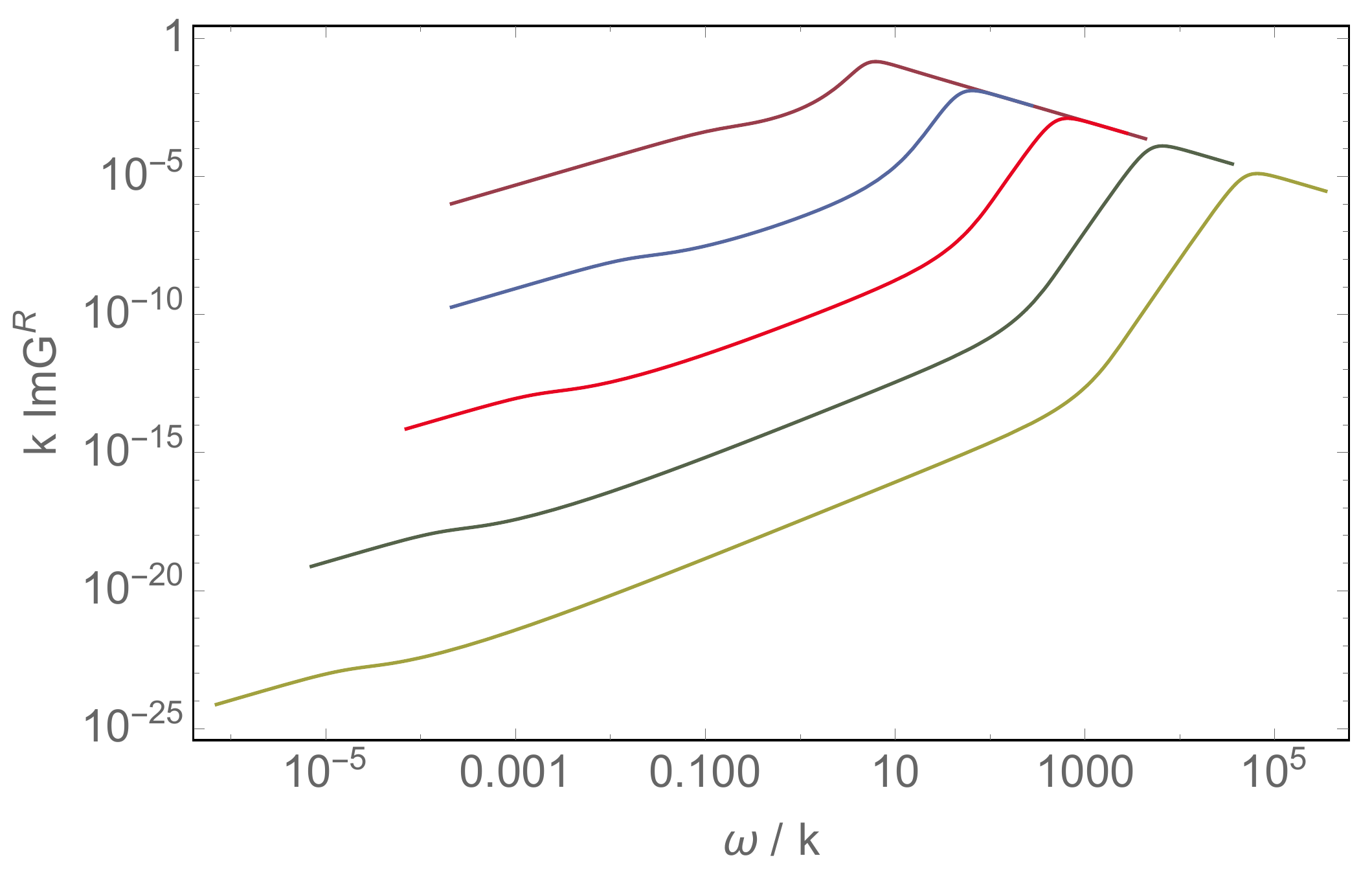}
\includegraphics[width=0.5\textwidth]{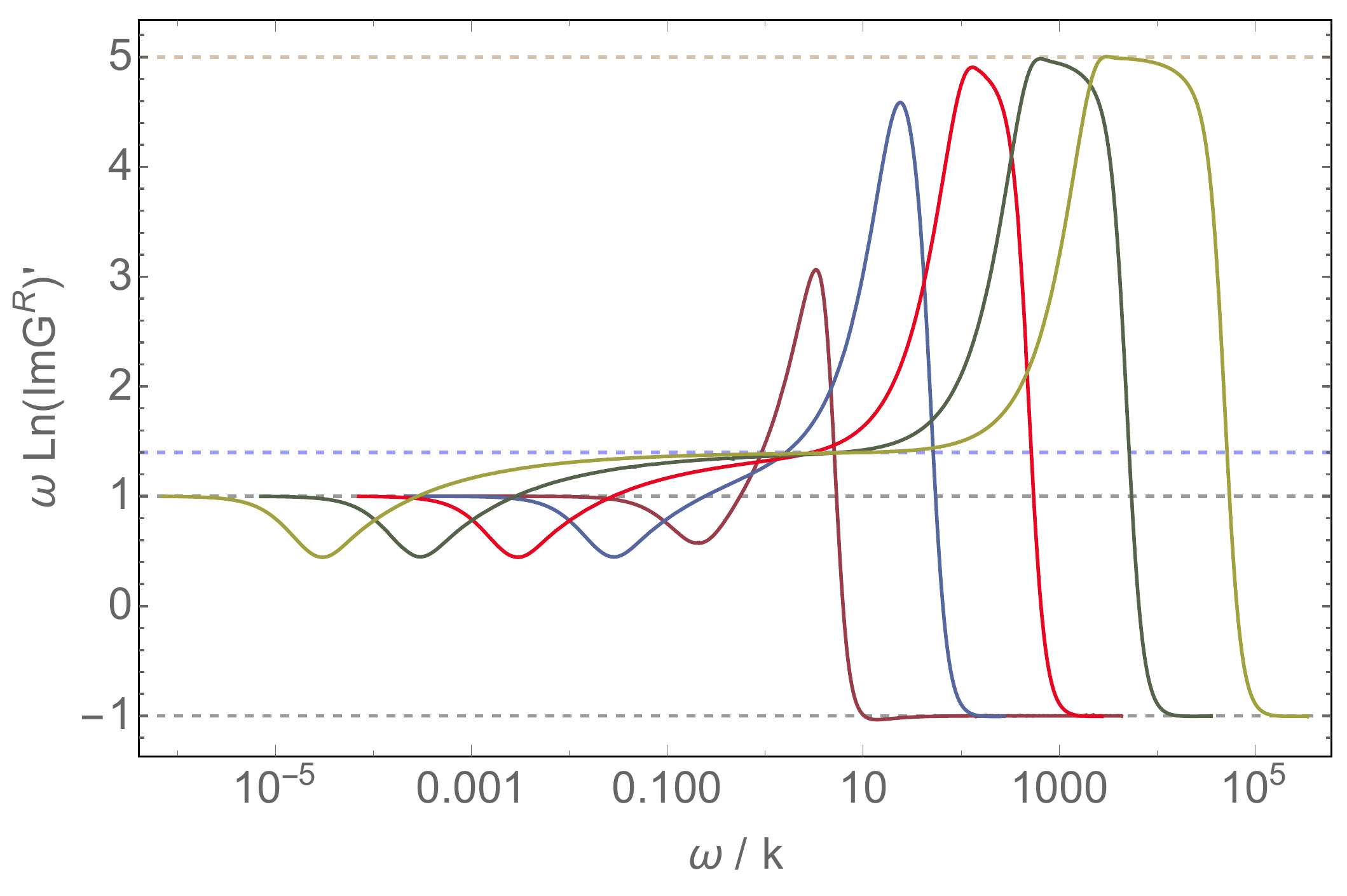}\\
\includegraphics[width=0.5\textwidth]{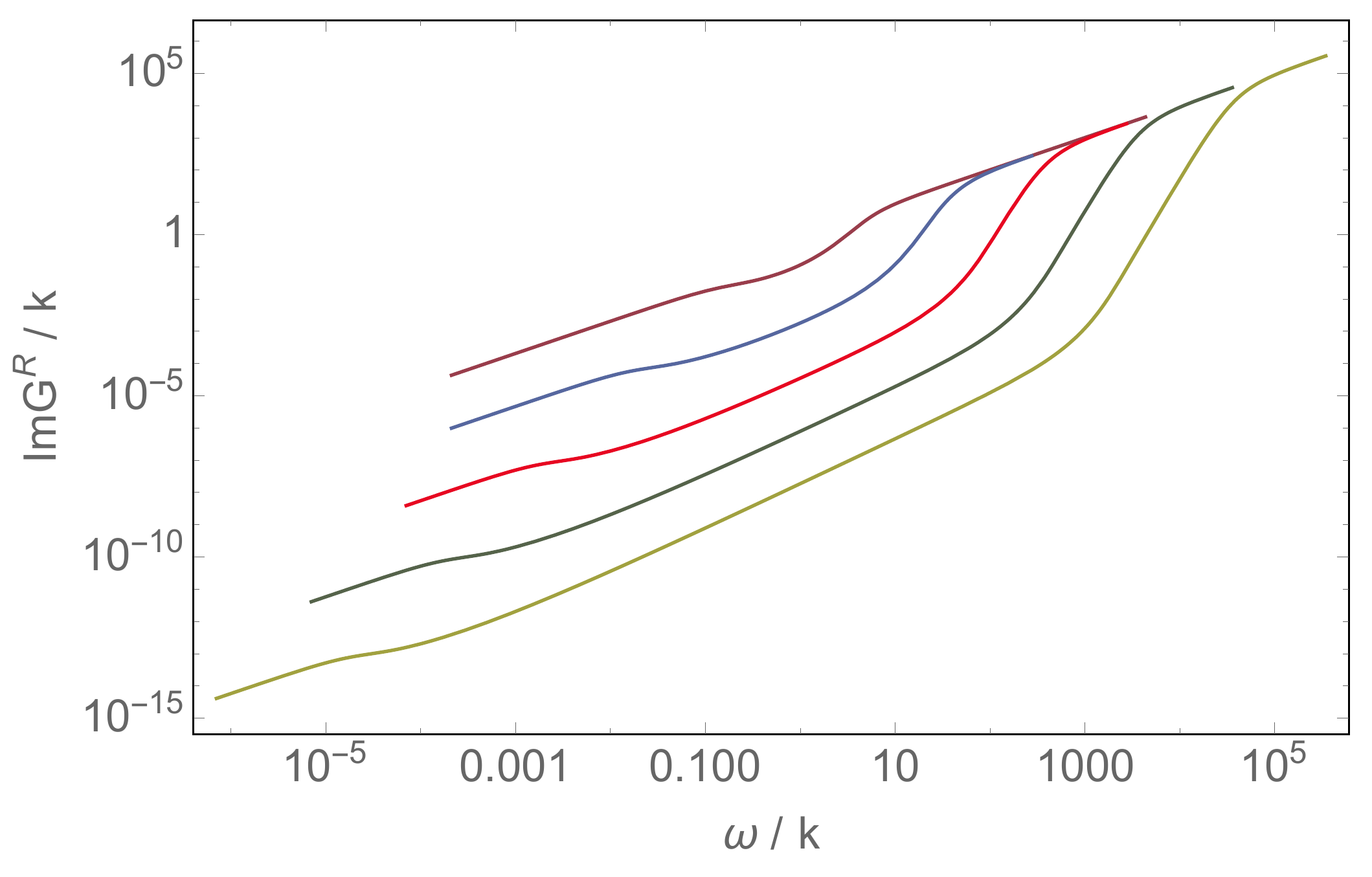}
\includegraphics[width=0.5\textwidth]{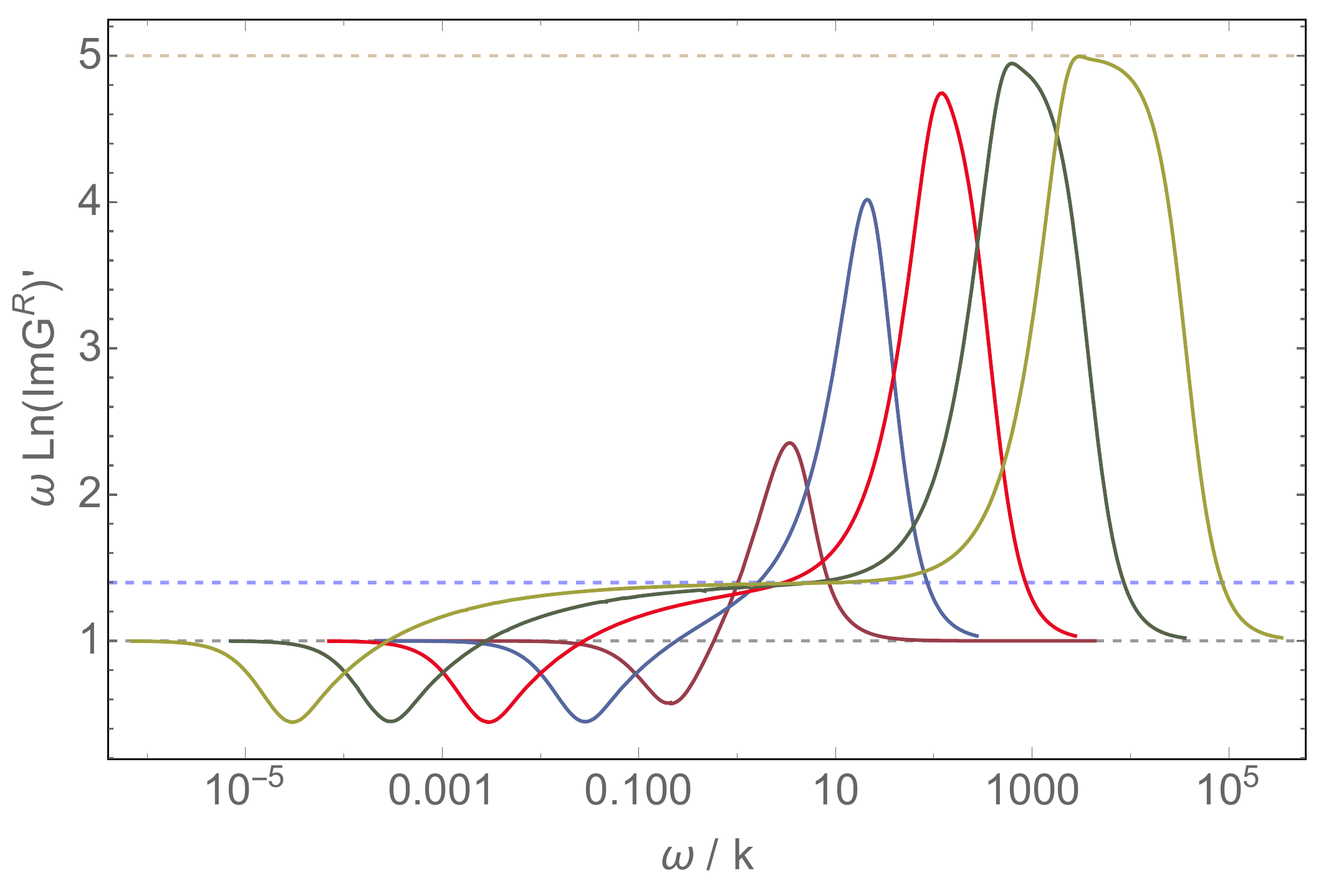}\\
%\center{\includegraphics[width=0.5\textwidth]{figs/plotsF3.pdf}}
\caption{Plots displaying the behaviour of the spectral function $\text{Im}{G}^R(\omega)$ for a scalar field with $\Delta=1$ (top plots) and
$\Delta=2$ (bottom plots) for various values of $\Gamma/k=0.7\times 10^n$: purple ($n=1$), blue ($n=2$), red ($n=3$), dark green ($n=4$), light green ($n=5$). 
In the left plot we see the build up of intermediate scaling regions as $\Gamma/k$ increases with scaling behaviour governed by 
the spectral functions $\text{Im}\mathcal{G}^R_{I}(\omega)\sim\omega^{7/5}$ and 
$\text{Im}\mathcal{G}^R_{II}(\omega)\sim\omega^5 $ of the two hyperscaling violation geometries. The right plot shows
the more stringent test of scaling behaviour by plotting the derivative of the logarithm.
\label{fig6}}
\end{figure}

We can now check these expectations by numerically constructing the spectral function $\text{Im}{G}^R(\omega)$ of the full boomerang RG flow. 
A key technical point in solving \eqref{radeq},
is that it is helpful to pull out an overall factor of $e^{-i\omega z}$ for $\psi$, where $z$ is defined in \eqref{zeddef}. Indeed we find
that this deals with the rapid oscillations of $\psi$ throughout the whole of the flow, including the
intermediate scaling regimes.
Our results, which involved considerable numerical effort, are presented in figure \ref{fig6} for various $\Gamma/k$. In particular, for the largest value of $\Gamma/k\sim 0.7\times 10^5$
we see the spectral function exhibits the intermediate scaling behaviour exactly as predicted above.

An important point to emphasise in the above analysis is that in each of the intermediate scaling regimes \eqref{fisr},\eqref{sndis} the mass term appearing in the Schr\"odinger potential \eqref{eq:sch_potential} 
is sub-dominant compared to the other term\footnote{As an aside we note that when considering the hyperscaling violation solutions \eqref{fisr} and \eqref{sndis} as UV complete solutions in themselves, one finds that the Schrodinger potential admits negative energy bound states, when $m^2<0$ and hence implies that the solutions are unstable for such scalars. However, as mentioned, this does not affect our conclusion concerning the intermediate scaling of the spectral functions on the boomerang RG flows.}. 
This means that the nature of the intermediate scaling that is displayed in figure \ref{fig6} will be essentially the same for all scalar modes which have a simple mass term, provided that
$m^{2}$ is much smaller than the second term inside the outer brackets of \eqref{eq:sch_potential} when evaluated in the intermediate regions. This implies an interesting type of universality for
the intermediate scaling behaviour of a wide class of operators, irrespective of their UV scaling dimensions, up to some maximum bound 
set by \eqref{eq:sch_potential}. This is analogous to  the universal scaling behaviour seen in standard RG flows in the far IR as $\omega\to 0$.

Furthermore, similar comments apply to scalar modes with different couplings to the background fields. For example, to illustrate the impact of different couplings, consider replacing the constant $m^2$ in \eqref{kg} with an $r$-dependent term $m^2(r)$. If $m^2(r)$ is still sub-dominant to
the other term in the Schr\"odinger potential there will be the same kind of universality in the intermediate scaling behaviour.
Alternatively, it may be possible to have top-down couplings in which $m^2(r)$ is the dominant term in an
intermediate scaling region which would again lead to intermediate scaling of the spectral function, but not
with the same kind of universality.

To conclude this section, we would like to highlight one more interesting feature of the spectral functions displayed in figure \ref{fig6}, independent of
the intermediate scaling. For both the $\Delta=2$ quantisation and the $\Delta=1$ quantisation we have 
$\text{Im}{G}^R(\omega)\propto\omega$ as $\omega\to 0$ when $\Gamma/k\ne 0$. Indeed this is an example of the standard universal scaling in the far IR of RG flows that we mentioned above. 
Now, for the $\Delta=2$ quantisation we also have 
$\text{Im}{G}^R(\omega)\propto\omega$ as $\omega\to\infty$. For this case, as the lattice deformation
is switched off, $\Gamma/k\to 0$, we continuously approach the $AdS_4$ result $\text{Im}{G}^R(\omega)\sim \omega$ for all $\omega$. On the other hand for 
$\Delta=1$ quantisation, we have $\text{Im}{G}^R(\omega)\propto\omega$ as $\omega\to 0$, but 
$\text{Im}{G}^R(\omega)\propto\omega^{-1}$ as $\omega\to\infty$. This implies that $\text{Im}{G}^R(\omega)$ has a maximum for some value of
$\omega$, as we see in figure \ref{fig6}. Furthermore, as $\Gamma/k\to 0$ this peak gets pushed closer and closer to $\omega\to 0$ and we do not
continuously approach the $AdS_4$ result of $\text{Im}{G}^R(\omega)\sim \omega^{-1}$ for all $\omega$. 
It would be interesting to study this feature in more detail.

\section{Final Comments}\label{fincom}

In this paper we have constructed a novel class of RG flows of $N=2$ STU gauged supergravity theory
that can be uplifted on the seven sphere to obtain solutions of $D=11$ supergravity. The solutions break translations, 
periodically, in both spatial
directions. The solutions
flow from $AdS_4$ in the UV to the same $AdS_4$ in the IR and on the way to the IR, for large enough deformations, 
they approach two distinct intermediate scaling regimes with hyperscaling violation.
It would be interesting to understand these novel RG flows directly from the dual field theory\footnote{Of course, here we are implicitly
assuming that if there are any other RG solutions of $D=11$ supergravity with the same asymptotic boundary deformations
then the ones we have constructed have the smallest free energy. It would be interesting to examine this issue in more detail: an analogous investigation at finite charge density was initiated in \cite{Donos:2011ut}.}.
In this context
the RG flows are driven by deformations of certain scalar and fermion bilinear operators of the dual CFT, with a specific periodic
dependence on the spatial coordinates governed by a single wavenumber. The intermediate scaling that we have seen is
associated with a class of deformations of the dual CFT within the framework of  a Q-lattice construction. It would be interesting to determine whether this behaviour persists for
more general deformations, by solving the associated partial differential equations.

We also constructed some finite temperature black holes solutions which lead to the RG flows in the $T\to 0$ limit.
As we explained, these black hole solutions are associated with a 
thermodynamic ensemble of the dual field theory in which we hold fixed the deformation parameters of the
pseudoscalar operators and the expectation values of the scalar operators. It is for this particular ensemble
that we are able to construct the black hole solutions by solving a system of ODEs. It would be interesting to
construct solutions in which the deformations of both sets of operators are held fixed, but to do this one
will have to consider a more general ansatz and solve a system of partial differential equations. At this stage it is not clear
to us whether the intermediate scaling that we have observed at $T=0$ will persist for finite $T$ in this other ensemble.

We have shown that for large enough deformations the spectral functions 
of certain scalar operators also exhibit scaling behaviour that is 
associated with the two intermediate scaling regimes, for certain intermediate values of frequency. Moreover,
this intermediate scaling behaviour is independent of the mass of the bulk scalar field and hence
independent of the conformal dimension of the scalar operator in the dual field theory.
Another interesting feature is that the scaling of the spectral function governed by the hyperscaling violation solution
can exist for a certain range of intermediate frequencies, even if the hyperscaling violation solution exhibits unstable behaviour for other frequencies.

It would be interesting to extend these investigations and calculate the thermoelectric conductivity for the solutions we have constructed.
As usual this involves analysing perturbations of the metric and gauge-fields about the solutions with prescribed boundary conditions. 
However, there is an intricate coupling between the gauge-fields and the scalar and pseudoscalar fields, parametrised
by the matrix $\mathcal{M}$ in \eqref{eq:Lag}, and as a consequence it will be a somewhat involved task to
calculate the conductivities, unlike for other Q-lattices. 
In general, the thermoelectric DC conductivity can be obtained by solving Navier-Stokes
equations on the black hole horizon \cite{Donos:2015gia}. 
For certain Q-lattice constructions these equations can be solved explicitly in terms of the horizon data \cite{Donos:2014uba,Donos:2014cya,Banks:2015wha}. 
Here, however, due to the coupling $\mathcal{M}$ it appears  
that this will not be the case and one will need to solve partial differential
equations on the horizon.

We have argued that at least for Q-lattice constructions which involve relevant operators governed by a dimensionless
parameter $\Gamma/k$, the appearance of a Poincar\'e invariant intermediate scaling regime should appear for large values
of $\Gamma/k$. For example, using \eqref{eq:tLag2} we can construct anisotropic Q-lattices using just the fields
$\lambda_2$, $\sigma_2$ with $\lambda_3=\sigma_3=0$. Although we have not checked the details, it seems very likely
that there will be an intermediate scaling regime governed by the same $k=0$ flow that we discussed in section \ref{keqzf}.
Furthermore, it seems unlikely that this Q-lattice construction will have a second intermediate scaling regime. More generally, it
is possible to make similar constructions in which the intermediate scaling regime is governed by an $AdS$ fixed point \cite{toapp}.

The Q-lattice constructions of this paper used a very specific global symmetry of the maximally supersymmetric
$N=8$ gauged supergravity theory. The 70 scalars of this theory parametrise the coset $E_{7(7)}/SU(8)$ and we utilised
a specific truncation that kept scalars parametrising two $SL(2)/SO(2)$ factors in $E_{7(7)}/SU(8)$. Furthermore, we exploited the
fact that the scalar potential was invariant under $SO(2)^2$ and this was utilised to construct
our Q-lattice ansatz. There are clearly many more $Q$-lattice constructions that could be made in the $N=8$ theory and it would be interesting 
to explore their properties. For example, one specific avenue is to utilise the consistent truncations\footnote{These truncations were used
in \cite{Bobev:2013yra} to construct supersymmetric Janus solutions. While the underlying physical setup is different, it would be interesting
to investigate whether there is any relationship with Q-lattice constructions of boomerang RG flows in some putative limit.}  that keep a single $SL(2)/SO(2)$ factor
that were discussed in \cite{Bobev:2013yra}. 

The solutions we have constructed all have vanishing gauge-fields and are associated with vanishing
charge density in the dual field theory. Some of the analogous type IIB anisotropic flows that we discussed in the introduction
have been generalised to finite charge density in \cite{Cheng:2014qia,Banks:2016fab} using a straightforward
consistent truncation. However, it is less clear how to add charge to the solutions that we have constructed with an isotropic metric
in the spatial directions of the field theory. In appendix \ref{appa}
we have 
identified a simple ansatz that is suitable for constructing charged solutions that are spatially anisotropic. Although our analysis has not been comprehensive, the constructions of some anisotropic RG flows that we have made did not reveal intermediate scaling behaviour. 
We think it would be worthwhile to investigate these charged solutions more systematically as well as looking for charged isotropic solutions.

\section*{Acknowledgements}
We thank Ofer Aharony, Nikolay Bobev, Gary Horowitz, Per Kraus, Don Marolf, Jorge Santos, Marika Taylor and David Tong for helpful discussions.
The work of JPG and CR is supported by the European Research Council under the European Union's Seventh Framework Programme (FP7/2007-2013), ERC Grant agreement ADG 339140. The work of JPG is also supported by STFC grant ST/L00044X/1, EPSRC grant EP/K034456/1, as a KIAS Scholar and as a Visiting Fellow at the Perimeter Institute.

\appendix
\section{STU gauged supergravity and some truncations}\label{appa}

The bosonic part of the Lagrangian for the $N=2$ STU gauged supergravity theory is given by
\begin{equation}\label{eq:Lag}
\mathcal{L} = R-\frac{1}{2}\sum_{i=1}^3\Big((\partial\lambda_i)^2 +\sinh ^2\lambda_i(\partial\sigma_i)^2\Big)
-\sum_a^4\mathrm{Re}\Big(F^{(a)+}_{\mu\nu}\mathcal{M}_{ab}F^{(b)+}\,^{\mu\nu}\Big)-\mathrm{g}^2 V\,,\end{equation}
where $ F^{(a)+}_{\mu\nu}=\frac{1}{2}(F^{(a)}_{\mu\nu}-i*F^{(a)}_{\mu\nu})$,
$V$ is the potential just depending on the scalar fields
\begin{equation}
V = -8\sum_i^3\cosh\lambda_i,
\end{equation}
and $\mathcal{M}$ is a rather complicated matrix that depends on both the modulus
and
the phase of the complex scalars $\Phi_i=\lambda_ie^{\sigma_i}$. 
An explicit expression can be found in \cite{Cvetic:1999xp,Cvetic:2000tb,Azizi:2016noi}.
It is clearly consistent with the equations of motion to set all of the gauge-fields to zero. It is also not difficult to
see that we can consistently set $\lambda_1=\sigma_1=0$. Finally we can set $\lambda_2=\lambda_3$ provided
that we have an ansatz in which $(\partial\sigma_2)^2=(\partial\sigma_3)^2$ and
this what we studied in the main part of this paper, and we also set $g^2=1/4$.

It is also of interest to look for simple frameworks in which we can have Q-lattice constructions that also
carry electric charge. We have found the following ansatz that can be used to construct 
RG flows with charge that depend on just one of the spatial directions. Introducing coordinates $(t,r,x,y)$ we can take
the only dependence on the $x$ direction to be via two of the periodic $\sigma_i$ fields, as in the ansatz: 
\begin{equation}\label{tone}
\lambda_1= \alpha,\qquad \lambda_2=\lambda_3= \gamma, \qquad \sigma_1=0,\qquad \sigma_2=-\sigma_3= kx
\end{equation}
and
\begin{equation}\label{tonet}
F^{(1)} = F^{(2)} = 0, \qquad F^{(3)} = F^{(4)} \equiv {F} \qquad \mathrm{with} \qquad {F}\wedge{F} = 0\,,
\end{equation}
The equations of motion of the STU theory can then be obtained from the Lagrangian
\begin{equation}\label{eq:tLag}
\mathcal{L} = R-\frac{1}{2}(\partial\alpha)^2-(\partial\gamma)^2-g^{xx}k^2\sinh^2\gamma-e^{-\alpha}{F}^2+8\,\mathrm{g}^2(\cosh\alpha+2\cosh\gamma).
\end{equation}
with all fields just depending on $(t,r,y)$.
One can now restrict to an electric ansatz for the fields in \eqref{eq:tLag} that just depend on the radial direction. For the scalar fields we take
$\gamma,\alpha$ to be functions of $r$ and for the gauge-field we take $F=d{A}$ with ${A}={A}_t(r)$.
A suitable anisotropic metric ansatz is given by $ds^2=Udt^2+U^{-1}dr^2+e^{2V_1}dx^2+e^{2V_2} dy^2$ with $U,V_1,V_2$ functions of $r$.

Finally, we point out that there is a closely related ansatz which leads to \eqref{eq:tLag} but with $\alpha\to -\alpha$. 
One way this can arise is by taking
$\lambda_1=\alpha$, $\lambda_2=\lambda_3=\gamma$, $\sigma_1=\pi$, $\sigma_2=\sigma_3=kx$, $F^{(1)}=F^{(2)}=0$ and
$F^{(3)}=-F^{(4)}\equiv F$.

%\bibliographystyle{utphys}
%\bibliography{helical}{}

\providecommand{\href}[2]{#2}\begingroup\raggedright\endgroup

\end{document}